\documentclass[fleqn,usenatbib]{mnras}

\usepackage{newtxtext,newtxmath}

\usepackage[T1]{fontenc}
\usepackage{float}
\usepackage{physics}
\usepackage{subfiles}
\usepackage{comment}
\usepackage{booktabs}
\usepackage{graphics}
\usepackage{comment}
\usepackage{graphicx}
\usepackage{hyperref}
\usepackage{xcolor}
\usepackage{lipsum}
\usepackage{tabularx}
\usepackage{graphicx}
\usepackage{orcidlink}
\usepackage{lipsum}
\usepackage{import}
\usepackage{graphics}
\usepackage{graphicx}
\usepackage{tabularx}
\usepackage{longtable}
\usepackage{tabularray}
\usepackage{multicol}
\usepackage{wrapfig}
\usepackage{supertabular}
\usepackage{graphicx}
\usepackage{multirow}
\usepackage{numprint}
\usepackage{siunitx}
\DeclareRobustCommand{\VAN}[3]{#2}
\let\VANthebibliography\thebibliography
\def\thebibliography{\DeclareRobustCommand{\VAN}[3]{##3}\VANthebibliography}

\usepackage{graphicx}	
\usepackage{amsmath}	
\usepackage{xcolor}

\definecolor{green_cust}{HTML}{00FF00}

\definecolor{blue_cust}{HTML}{0066ff}

\definecolor{gold_cust}{HTML}{FFBF00}

\definecolor{purple_cust}{HTML}{800080}

\definecolor{pink_cust}{HTML}{DC267F}

\definecolor{perfect_green}{HTML}{4FBF26}

\definecolor{crimson}{HTML}{DC143C}

\newcommand{\grb}{AT2023vfi}
\newcommand{\vfi}{AT2023vfi}

\newcommand{\gfo}{AT2017gfo}

   \title[Collisional data for Ce II-IV with nebular emission modelling]{Ce II-IV emission in kilonovae with R-matrix collision strengths and distorted wave recombination rates}

\author[Leo Patrick Mulholland et al.]{
Leo P. Mulholland$^{1}$\thanks{E-mail: lmulholland25@qub.ac.uk}\orcidlink{0009-0003-2668-5589}, Niamh Ferguson$^{2}$\orcidlink{0009-0005-3148-513X}, Luke J. Shingles$^{1}$\orcidlink{0000-0002-5738-1612}, Catherine A. Ramsbottom$^{3,1}$\orcidlink{0000-0003-1579-8556},  \newauthor{}~Connor P. Ballance$^{1}$\orcidlink{0000-0003-1693-1793},  and Stuart A. Sim$^{1}$\orcidlink{0000-0002-9774-1192}
\\
$^{1}$Astrophysics Research Centre, School of Mathematics \& Physics, Queens University Belfast, BT7 1NN, Northern Ireland. \\ 
$^{2}$Department of Physics, University of Strathclyde, 107 Rottenrow, G4 0NG, United Kingdom\\
$^{3}$ School of Physics, University College Dublin, Belfield, Dublin 4, Ireland}

\date{Accepted XXX. Received YYY; in original form ZZZ}

\pubyear{\the\year{}}

\begin{document}
\label{firstpage}
\pagerange{\pageref{firstpage}--\pageref{lastpage}}
\maketitle



\begin{abstract}
We provide transition probabilities and Maxwellian-averaged collision-strengths for the electron-impact excitation of Ce {\sc ii} - {\sc iv}. These ions are potentially abundant in kilonovae (KNe) and feature both forbidden and allowed transitions in the infrared. These atomic data are obtained using the {\sc grasp}$^0$ atomic structure code, interfaced with the non-perturbative {\sc darc} R-matrix close-coupling codes.  We additionally use {\sc autostructure} to produce dielectronic and radiative recombination rate coefficients for Ce {\sc ii} - {\sc vi}. Using this data, we calculate an approximate ionization balance including non-thermal ionization rates with deposition rates from realistic models. We find  Ce {\sc ii} - {\sc iv} to be of roughly equal abundance, subject to ejecta conditions. We find that generally we need $\gtrsim$ 10$^{-3}$ M$_{\odot}$ of Ce in total to produce emission of comparable luminosity to the features in the spectrum of AT2023vfi. In particular, it is found that [Ce {\sc iv}] 4.5 $\mu$m requires both strong ionization and weak recombination, as well as a production of Ce considerably stronger than that of the solar abundance in order to have luminosity comparable to the apparent emission in this wavelength vicinity in \grb.  
With such strict requirements on individual element masses, the detection of individual lanthanides on a line-by-line basis in the nebular phase seems unlikely - but lanthanides may remain important for cooling and early time analyses.
\end{abstract}

\setlength{\tabcolsep}{4pt}


\section{Introduction}

The observation of a binary-neutron-star merger in 2017 via the gravitational wave detection of GW170817, with corresponding electromagnetic signature AT2017gfo, has provided direct evidence of $r$-process nucleosynthesis \citep{watson2019identification,gillanders2024modelling,hotokezaka2023tellurium,Hotokezaka2022WSe,domoto2022lanthanide,tanaka2023cerium}. A recent study by \cite{domoto2022lanthanide} has highlighted the potential presence of La {\sc iii} and Ce {\sc iii} in the ejecta of AT2017gfo. These lanthanides lie towards the left-side of the periodic table, giving them relatively sparse electron structures leading to strong lines as argued by \cite{domoto2022lanthanide}. Lines of Ce {\sc ii} and {\sc iii} have also been found to contribute significantly to local-thermodynamic-equilibrium (LTE) simulations such as those featured in \cite{shingles2023self,gillanders2026improved}. Given KNe are thought to be sites of lanthanide production, Ce presents a particularly good case to study. It is expected to be among the more abundant of the lanthanides and has a relatively simple atomic structure compared with other lanthanides.

  There have been significant efforts to converge the opacity contribution from Ce and the works of \cite{kato2024systematic,tanaka2020systematic,quinet2020current,flors2026calibrated,gaigalas2024theoretical,carvajal2021multiconfiguration} have produced large atomic structure calculations from which expansion opacities can be calculated. Thus far there has been mixed agreement within these calculations. The impact of lanthanides in KNe spectra has been under some debate in the past years. In particular,  \cite{pognan2026lanthanide} argue they are unable to reproduce the seemingly blackbody continuum seen in the second spectrally observed kilonova AT2023vfi - which may lead to alternate interpretations. Additionally, the {\sc tardis} simulations of \cite{gillanders2026improved} indicate a factor of $\sim 20$ lower fraction of lanthanides is required to reproduce the rough spectrum in the (approximately) photospheric phases of AT2017gfo.

In addition to its known importance for opacity, Ce possesses allowed lines in the NIR. These include [Ce {\sc iii}] $\sim$ 1.6 $\mu$m and $3.0\mu$m. The former transition has been suggested as an absorption feature in AT2017gfo by \cite{domoto2022lanthanide} and will be further studied by Rahmouni et al. ({in prep}). The coincidence of forbidden [Ce {\sc iv}] $4.5\mu$m with excess flux in \vfi~has previously been pointed out by \cite{gillanders2025analysis} (see their table 3). Previous Non-LTE studies including Ce \citep{PognanNLTE} have made use of the semi-empirical formula of \cite{van1962rate} for allowed line collision strengths, and also the formulae of \cite{Axelrod1980} for forbidden line collision strengths and recombination rates.  In this work we provide new atomic data for the improvement of modelling Ce in KNe simulations. In particular, we provide Dirac $R$-matrix collision strengths for Ce {\sc ii} to {\sc iv}  and recombination rate coefficients using the {\sc autostructure} code \citep{badnell1986dielectronic, badnell}.

The remainder of this paper is structured as follows. In Section \ref{sec:atomic} we calculate compact yet representative atomic structures of Ce {\sc ii} to {\sc iv} using the {\sc grasp}$^0$ code. The energy levels are compared with literature values from the {\sc nist} \citep{nist} database. Where possible, transition probabilities are compared with literature values such as those of \cite{quinet2020current,flors2026calibrated}. In Section \ref{sec:rmatrix} we present details of the $R$-matrix collision strength calculations and collision strengths. In Section \ref{sec:recomb} we show the details of the recombination calculations using {\sc autostructure}. The rate coefficients of both dielectronic and radiative recombination are calculated for low-energy metastable levels. These, together with the collision strengths, shall be made available. In Section \ref{sec:crm} we combine the above data into a collisional radiative model. We first estimate the mass of specifically Ce {\sc iv} required to observe forbidden line emission in the optically thin limit. We then calculate an ionization balance, assumed to be dominated by fast ionizing electrons and thermal recombining electrons. The distribution of fast electrons and ionization rates are calculated by solving the Spencer-Fano equations \citep{spencer1954energy,kozma1992gamma} for which we use the {\sc pynonthermal} code of \cite{luke_shingles_2026_19814117}. We then estimate the amount of Ce required to produce observable nebular emission under KNe conditions. Finally in Section \ref{sec:conclusions} we present our conclusions and outlook for future work and observations.

\section{Atomic Structure} \label{sec:atomic}

\begin{table}
    \centering
    \begin{tabular}{c l }
    \hline \\

    Ce {\sc ii} Model  & 5s$^2$ 5p$^6$ 4f$^1$\{ 5d$^2$, 5d$^1$ 6s$^1$, 6s$^2$, 6s$^1$ 6p$^1$, 6p$^2$  \};           \\
     - 18 CSF          & 5s$^2$ 5p$^6$ 4f$^2$\{5d$^1$, 6s$^1$, 6p$^1$\};           \\
                       & 5s$^2$ 5p$^6$ 5d$^1$\{6s$^2$, 6p$^2$\};           \\
                       & 5s$^2$ 5p$^6$ 5d$^2$\{6s$^1$, 6p$^1$\};           \\
                       & 5s$^2$ 5p$^6$ \{ 4f$^3$, 5d$^3$, 6p$^3$\};           \\
                       & 5s$^2$ 5p$^6$ \{ 6s$^1$ 6p$^2$\};           \\
                       & 5s$^2$ 5p$^5$ 4f$^3$ 6s$^1$ ;           \\

\\
    Ce {\sc iii} Model  & 5s$^2$ 5p$^6$ \{4f$^2$, 5d$^2$\};           \\
     - 13 CSF          & 5s$^2$ 5p$^6$ 4f$^1$\{ 5d$^1$, 6s$^1$, 6p$^1$, 6d$^1$  \};  \\
                       & 5s$^2$ 5p$^6$ 5d$^1$\{ 6s$^1$, 6p$^1$, 6d$^1$  \};           \\
                       & 5s$^2$ 5p$^5$ \{4f$^3$,4f$^2$ 6p$^1$, 4f$^1$5d$^2$ \};           \\
                       & 5s$^2$ 5p$^4$ 4f$^4$;           \\
\\
    Ce {\sc iv} Model  & 5s$^2$ 5p$^6$\{4f$^1$, 5d$^1$, 5f$^1$, 6s$^1$, 6p$^1$, 6d$^1$, 7s$^1$, 7p$^1$, 7d$^1$, 8s$^1$\};           \\
     - 24 CSF          & 5s$^2$ 5p$^5$ 4f$^1$\{5d$^1$, 5f$^1$, 6s$^1$, 6p$^1$ \};           \\
                       & 5s$^2$ 5p$^5$ \{4f$^2$, 5f$^2$, 5d$^2$, 6d$^2$, 7s$^2$ \};           \\
                       & 5s$^2$ 5p$^5$ 6p$^1$ 7p$^1$;           \\
                       & 5s$^2$ 5p$^4$ 4f$^3$;           \\
                       & 5s$^1$ 5p$^6$ 4f$^1$ \{ 5d$^1$,6s$^1$\};           \\
                       & 5s$^0$ 5p$^6$ 4f$^3$ ;           \\

    \hline\\
    \end{tabular}
    \caption{Electronic configurations for the atomic structure models.}
    \label{tab:csfs}
\end{table}

\begin{figure*}
    \centering
    \includegraphics[width = \linewidth]{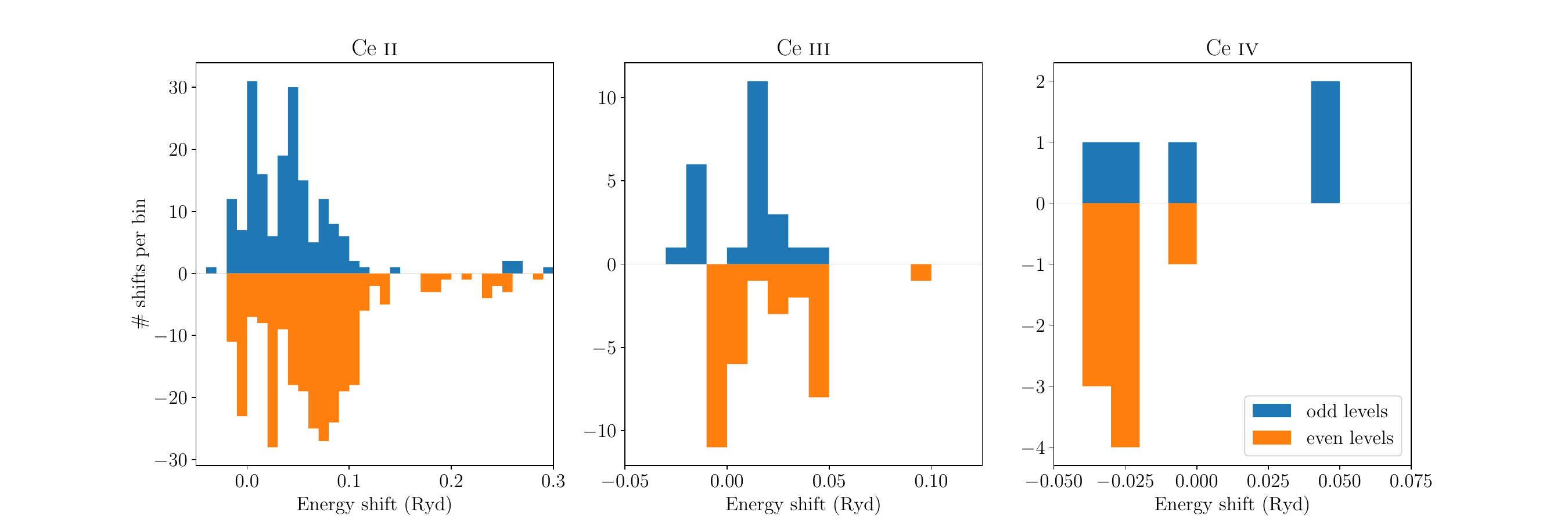}
    \caption{The distribution of energy differences $E_{\text{NIST}}-E_{\text{theory}}$ for each of the three models calculated in this work. The histograms are calculated with a bin width of $0.01$ Ryd. Positive bars are the number of shifts per bin for the odd levels, and negative counts are for even levels.}
    \label{fig:energycomp}
\end{figure*}

The target wave functions of the Ce ions studied here are obtained using an updated variant {\sc grasp}$^0$ of the MCDF atomic structure code \citep{Grant80}.  Within the configuration-interaction space, the Hamiltonian, 
\begin{equation}
H^{N} = \sum_i \bigg(c \boldsymbol{\alpha}\cdot\boldsymbol{p}_i + (\beta -I_4)c^2 -\frac{Z}{r_i}\bigg) + \sum_{i>j}\frac{1}{r_{ij}} \label{eq:ham},
\end{equation}
is diagonalized. The radial functions are optimized by the extended-average-level method. In Eq. \eqref{eq:ham}, $\boldsymbol{\alpha}$ and $\beta$ are the set of four Dirac-matrices, $c$ is the speed of light, $r_i$ is the radial position of electron $i$, $\boldsymbol{p}_i$ is the momentum-operator of electron $i$, $r_{ij}$ is the inter-electron distance, $Z$ is the nuclear charge and $I_4$ is the 4 $\times $ 4 identity matrix. 
We also calculate the Einstein A-coefficients of spontaneous emission. For electric-dipole transitions, these are given by,
\begin{equation}
	A_{j \to i } = \frac{16\pi^3}{3h\epsilon_0\lambda^3 g_j} | \bra{\psi_j} P \ket{\psi_i} |^2,
\end{equation}
where $P$ is the electric-dipole operator, $\lambda$ the transition wavelength and  $\psi_{i}$ are the eigenstates of Eq. \eqref{eq:ham} with statistical weights $g_{i} = 2J_{i}+1$. For this work, these values are corrected according to a shift to spectroscopic wavelengths with,
\begin{equation}
	A^{\text{calib.}}_{j \to i } = \frac{\lambda_{\text{calc.}}^3}{\lambda_{\text{expt.}}^3} A^{\text{theor.}}_{j \to i }. \label{eq:shifting}
\end{equation}
Higher order multipole transition probabilities are similarly formulated, and are corrected with increasing powers of the wavelength ratio. Recently, a mass production of data was performed by \cite{flors2026calibrated,da2025systematic} using optimization of fictitious mean configurations with the {\sc fac} code \citep{gu2008flexible}. We compare our A-values with this source where possible. Additionally, a large set of calculations were performed by \cite{carvajal2021multiconfiguration}. In the following calculations, we used the configuration expansions given in Table \ref{tab:csfs}, with the order of orbital optimization detailed in Appendix \ref{sec:graspstructure}. As a general note, it is required for these calculations to keep the configuration space relatively small to avoid the subsequent collision calculations (see Section \ref{sec:rmatrix}) becoming unmanageably large. The {\sc grasp}$^0$ code was modified to produce outputs compatible with the {\sc jj2lsj} code of \cite{gaigalas2017jj2lsj}, which was used to obtain $LS$ compositions to compare with literature designations. This allowed for the comparison to standard energy values and energy shifting. On Figure \ref{fig:energycomp}, we show the distribution of energy shifts. We now give details of the individual ion calculations.

\begin{figure*}
    \centering
    \includegraphics[width = \linewidth]{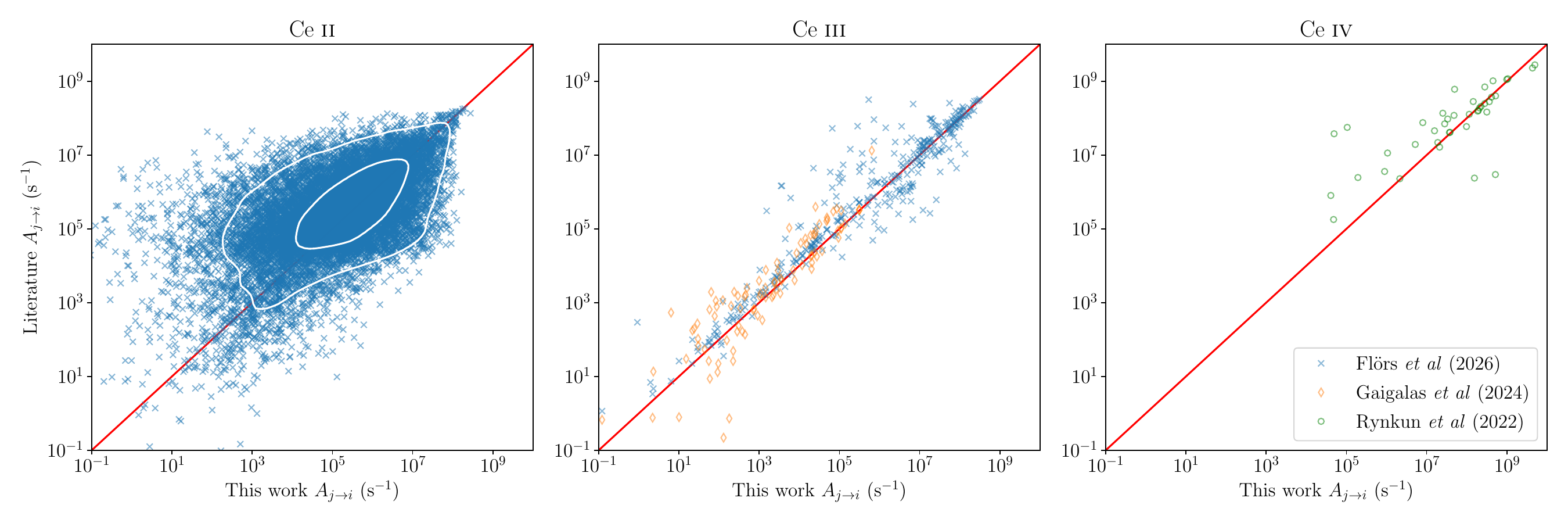}
    \caption{A-value (s$^{-1}$) comparison between the {\sc grasp}$^0$ models presented here and other theoretical calculations. Due to the number of lines in Ce {\sc ii}, we only compare with one other calibrated dataset of \citet{flors2026calibrated}. For this same reason, we also draw contour lines encompassing 39.3\% and 86.5\% of the data points. For Ce {\sc iii} we additionally compare with \citet{gaigalas2024theoretical}. Finally, for Ce {\sc iv} we compare with \citet{rynkun2022theoretical}.}
    \label{fig:avalue_comp}
\end{figure*}

\subsection{Ce {\sc ii}}

Ce {\sc ii} represents a complex atomic system, with the ground configuration 
4f$^1$\,5d$^2$. The coupling of multiple f and d electrons makes for a dense level structure. While the {\sc nist} database lists 491 levels, only a small proportion of  these have identified $LSJ$ labels. In addition, the strong configuration mixing renders many of the designations difficult to match between calculations.

For KNe applications we aim to optimally represent the target structure in the low-temperature nebular phase. Hence all levels below $\sim$ 0.55 Ryd ($\approx 7$ eV) were shifted to their exact experimental positions. This energy range roughly covers the accessible levels in nebular KNe. The two configurations 4f$^1$\,5d$^1$\,6p$^1$ and 4f$^1$\,6s$^1$\,6p$^1$ exhibited significant configuration mixing within themselves and additionally with each other. The typical $LSJ$ purity of these levels is around $\sim 30$\%. For this reason, such levels were shifted to the closest energy match with the {\sc nist} database in the absence of matches in their quantum numbers. We additionally shifted levels $\gtrsim 0.55$ Ryd where the $LSJ$ purity posed no ambiguity when matching to literature designations ($\gtrsim 70$\%).

 In the case of Ce {\sc ii} (left panel of Figure \ref{fig:energycomp}), the odd and even levels are roughly equally represented with mean absolute shifts of $\sim0.07$ Ry. The biggest discrepancies in the low lying energies occurred in the 4f$^1$\,5d$^1$\,6s$^1$ and $4$f$^3$ configurations, both are poorly represented in this model and likely require much more electronic correlation. These levels produce the shifts $> 0.2$ Ryd. As this is a highly correlated system, it is perhaps expected that more configurations than those included are required for higher accuracy energies.  
Given  the requirement of a reduced configuration set for the $R$-matrix calculation, the correct ground state and roughly equal accuracy of the odd and even levels gives confidence in the structure quality for this purpose - particularly for a case such as this where we are able to correct each of the important level energies.

To compare with literature data, we compare Einstein A-values with those from \cite{flors2026calibrated}. For simplicity, only those transitions with experimental wavelengths from each pair of  calculations are compared. This is shown in the left panel in Figure \ref{fig:avalue_comp}. Due to the large number of allowed lines, the comparison plot is quite densely populated. As a visual aid we include contour lines encompassing $39.3$\% and $86.5$\% of the data. It is clear that the second contour is scattered closely to the line of equality - within 1-2 dex. For a system as highly correlated as Ce {\sc ii}, there is relatively good agreement. In particular,  the calculations are at the level of agreement typically seen in \cite{flors2026calibrated} when comparing with other calculations.

\subsection{Ce {\sc iii}}
With a ground-state of 4f$^2$ with a closely excited 4f$^1$\,5d$^1$  configuration,  Ce {\sc iii} presents an example of a relatively simple lanthanide. Recent studies of Ce {\sc iii} include the works of the Mons and Lithuania groups \citep{carvajal2021multiconfiguration,gaigalas2024theoretical} and \cite{fischerCe2+}.  

The distribution of the present calculated energy levels compared with the {\sc nist} standard values is shown in the middle panel of Figure \ref{fig:energycomp}. Generally, the low-lying states are well represented. The odd and even states have an average absolute error of $\sim 0.02$ Ryd.  Notably the largest error occurs for the 4f$^2$ $^1$S$_0$ state, this is also the case for literature calculations \citep{carvajal2021multiconfiguration,gaigalas2024theoretical} and is due to the relatively few ways to construct a $^1$S$_0$ state to mix with this level.

The {\sc nist} database lists no transition probabilities for this ion. For this reason, we compare with literature theoretical datasets. To date, the largest calculation with a purely ab-initio approach is that of \cite{gaigalas2024theoretical} using the {\sc grasp2018} code. Since Ce {\sc iii} has robust $LSJ$ identifications, we are able to compare directly with this calculation despite their use of ab-initio wavelengths. On the middle panel of Figure \ref{fig:avalue_comp}, we compare our calculated values with this dataset as well as those of \cite{flors2026calibrated}. Only electric-dipole transitions are compared as these are the only available lines in each of the literature datasets. It can be seen that there is generally good agreement between the three datasets. Arguably, the most trustworthy dataset is that of \cite{gaigalas2024theoretical}, who includes QED corrections and large configuration expansions. Since the calculations of \cite{gaigalas2024theoretical} are not calibrated, the disagreement for weaker transitions may be explained by their longer wavelengths and therefore high sensitivity to the correction in Eq. \eqref{eq:shifting}. Nevertheless, there is reasonable agreement to within $\lesssim 1$ dex for the majority of the transitions considered. While the wavefunctions here will typically be less converged than particularly those in \cite{gaigalas2024theoretical} - the agreement of the A-values in the middle panel of Figure \ref{fig:avalue_comp} gives confidence in the quality of the target wavefunctions adopted in the subsequent $R$-matrix collision calculations.

\subsection{Ce {\sc iv}}

Ce {\sc iv} has ground configuration of 5p$^6$\,4f$^1$. The high effective charge of this system allows for the promotion of 5p electrons to lie beneath the ionization potential, such as the configurations 5p$^5$\,4f$^2$ and 5p$^5$\,4f$^1$\,5d$^1$. There are few spectroscopic studies of such configurations, although there are some theoretical works such as \cite{rynkun2022theoretical}. For this work, the most interesting aspect of Ce {\sc iv} is the low lying forbidden transition 5p$^6$\,4f$^1$ $^2$F$^o_{7/2}$ $\to$ $^2$F$^o_{5/2}$, where the emitted photon lies in the MIR at $\sim 4.4 \mu m$ and is coincident with and apparent emission line in AT2023vfi \citep{levan2024heavy,gillanders2025analysis}, and photometry in AT2017gfo \citep{villar2018spitzer}. The modest ionization state of the ejecta due to $\beta$-decay electrons makes this ion potentially viable in the ejecta \citep[see studies such as][]{pognan_steady,PognanNLTE}. 

 Levels with single excitations, of the form 5p$^6$ $nl^1$, were shifted where possible, with an average shift of $\sim 0.03$ Ryd. Because of the low purity of the  5p$^5$\,$nl^1 n'l'^1$ configurations ($\sim 30-50$\%), these are difficult to match between calculations and experimentally verify. These levels were kept at their \emph{ab initio} values. We compare only transitions involving these configurations with \cite{rynkun2022theoretical}. This is shown on the rightmost panel of Figure \ref{fig:avalue_comp}. Good agreement is seen for a few transitions involving the 5p$^6$\,7p$^1$ levels, noting these states are heavily mixed with the 5p$^5$ configurations. Additionally, there are no spectroscopic energies for these states available in {\sc nist} and therefore the transition probabilities may be quite susceptible to the energy correction in Eq. \eqref{eq:shifting}.

\section{Electron-Impact-Excitation} \label{sec:rmatrix}

To calculate the collision strengths required for modelling, we employ the Dirac-R-matrix codes ({\sc darc}; \citealt{chang1975r,Ballance2026_DARC}), which adopt a close-coupling-formulation \citep{burke2011r}. We do not reproduce the theory in full here and only provide the calculation parameters. For a more rigorous account of the theory we refer the interested reader to \cite{burke2011r}. 

Essentially, $N_{\text{cc}}$ target states  (calculated in Section \ref{sec:atomic}) are coupled to the  continuum to form a total symmetry $J\pi$, or partial wave, by means of a close-coupling expansion. The continuum wavefunctions are expressed as a basis set expansion of $n_c$ radial functions
subject to boundary conditions at $r=a$.
This boundary defines the inner and outer regions of our configuration space.
 For $r<a$, interactions between the target and the incident electron are considered strong and therefore electronic exchange and correlation are explicitly calculated. For $r>a$, the scattering reduces to essentially a two body problem where exchange is ignored. These two solutions are matched at the boundary by the $R$-matrix,
\begin{equation}
    R^{J\pi}_{ij}(E)=\frac{1}{2a}\sum_{k} \frac{w_{ik}(a)w_{jk}(a)}{E^{N+1}_{k} -E} \label{eq:rmatrix},
\end{equation}
where $w_{ik}$ are the surface amplitudes of the continuum electron at $r=a$ and the summation $k$ runs over the eigenvalues ($E^{N+1}$) of the of the $N+1$ Hamiltonian ($H^{N+1}$). From the $R$-matrix solutions, energy-dependent collision strengths $\Omega_{ij}$ for a transition induced between levels $i$ and $j$ are obtained. These collision strengths are related to the cross section by,
\begin{equation}
\sigma_{i\to j} = \pi a_0^2 \left(\frac{I_H}{g_i\epsilon_i}\right) \Omega_{ij},
\end{equation}
where $I_H$ is the ionization potential of atomic hydrogen and $\epsilon_i$ is the incident (pre-collision) energy of the continuum electron.

Typically for excitation in astrophysics, modelling assumes a thermal electron distribution (although there are additionally non-thermal contributions to all processes, including excitation e.g \citealt{kozma1992gamma}). For such thermal gases, one employs the Maxwellian-averaged collision strength (also known as the effective collision strength) given by,

\begin{equation}
    \Upsilon_{ij} = \int_0^{\infty}  e^{- \epsilon_j /kT_e} \Omega_{ij}   {\rm d} \left( \frac {\epsilon_j} {kT_e}\right) , \label{eq:eff}
\end{equation}
where $\epsilon_j$ is the scattered (post-collision) energy of the continuum electron, $k$ is the Boltzmann constant and $T_e$ is the thermal electron temperature. This is directly related to the rate coefficients of collisional excitation and de-excitation by, 
\begin{align}
       q_{i\to j}(T_e) &= \frac{4}{\sqrt{2\pi m_e}} \frac{\pi a_0^2 I_H}{g_i} \frac{1}{(kT_e)^{1/2}} \Upsilon_{ij}(T_e) e^{-E_{ij}/kT_e} \hspace{1mm}\label{eq:rates}, \\
    q_{j \to i}(T_e) &= \frac{g_i}{g_j} e^{E_{ij} /kT_e} q_{i\to j},
\end{align}
where $m_e$ is the electron mass and $E_{ij}$ is the transition energy. In the remainder of this section, we describe the calculation details and show sample collision strengths for the respective ions.
\begin{table}
    \centering
    \begin{tabular}{cccccccccc}
    \toprule
                & $n_c$ & $a$ & $N_{\text{cc}}$ & dim($H^{N+1}$) & $\#\epsilon_j$ &$\delta\epsilon_j$& $\#J\pi$   & Topup\\
\midrule
   Ce {\sc ii}  &  11   &  17.94  & 536 & 50208 & 31632 & 9.16$\times 10^{-5}$ &96&\checkmark\\
   Ce {\sc iii} &  20   &  17.94  & 100 & 13080 & 32512& 9.06$\times 10^{-5}$  &96&\checkmark\\
   Ce {\sc iv}  &  25   &   24.22   & 130 & 25200 & 32768 & 6.10$\times 10^{-5}$ &64&\checkmark\\
    \bottomrule     
    \end{tabular}
    \caption{Parameters used in the {\sc darc} calculations.}
    \label{tab:rmatrixParameters}
\end{table}

\begin{figure*}
    \centering
    \includegraphics[width = 0.7\linewidth]{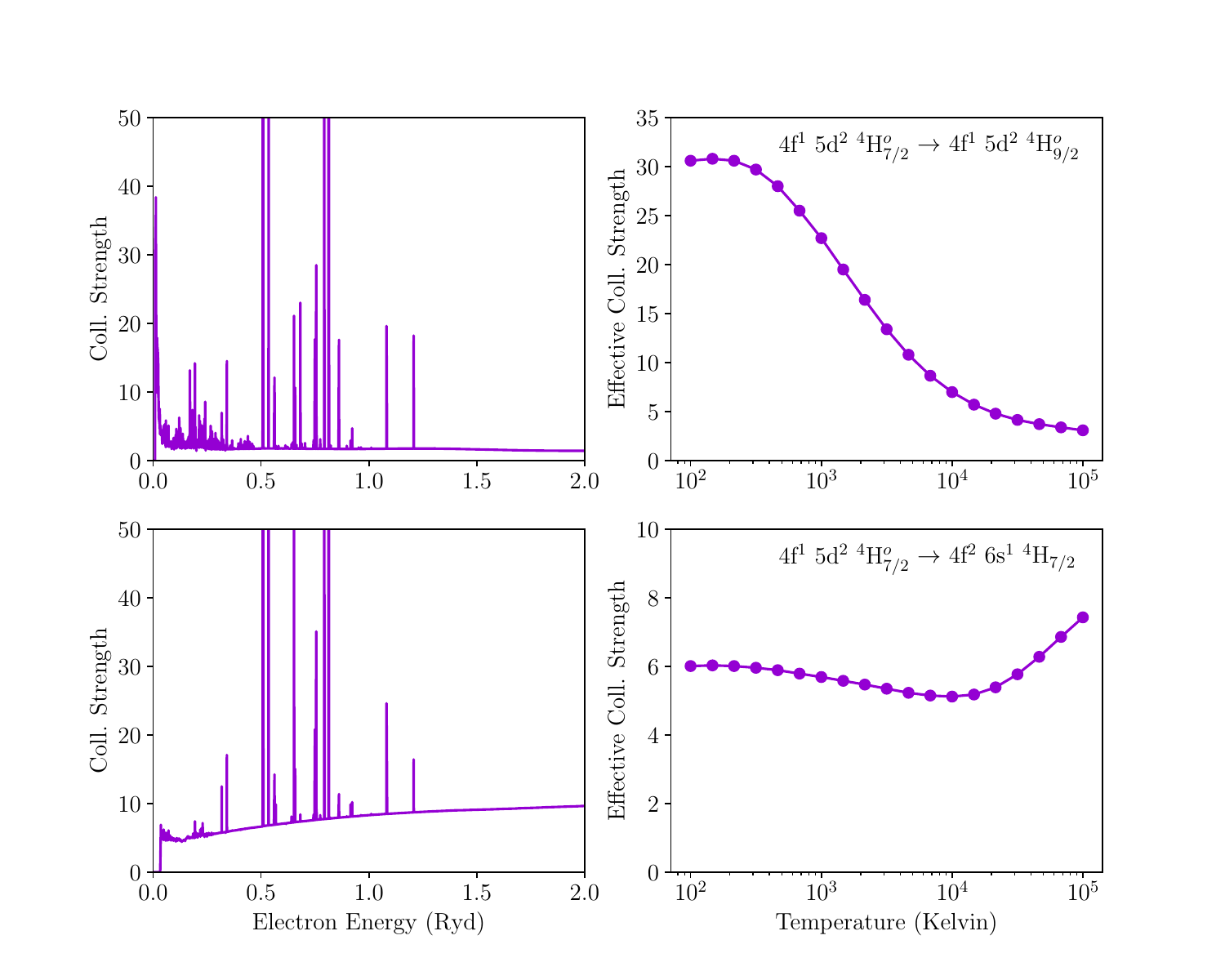}
    \caption{Collision strengths for two transitions of Ce {\sc ii}. The top two panels show the collision strengths and effective collision strengths for the forbidden transition 4f$^1$ 5d$^2$ $^4$H$_{7/2}^o$ $\to$ 4f$^1$ 5d$^2$ $^4$H$_{9/2}^o$ (Level 1 $\to 2$). The bottom two panels show the allowed transition 4f$^1$ 5d$^2$ $^4$H$_{7/2}^o$ $\to$ 4f$^2$ 6s$^1$ $^4$H$_{7/2}$ (Level 1 $\to $19).}
    \label{fig:CeIICollisions}
\end{figure*}

\begin{figure*}
    \centering
    \includegraphics[width = 0.7\linewidth]{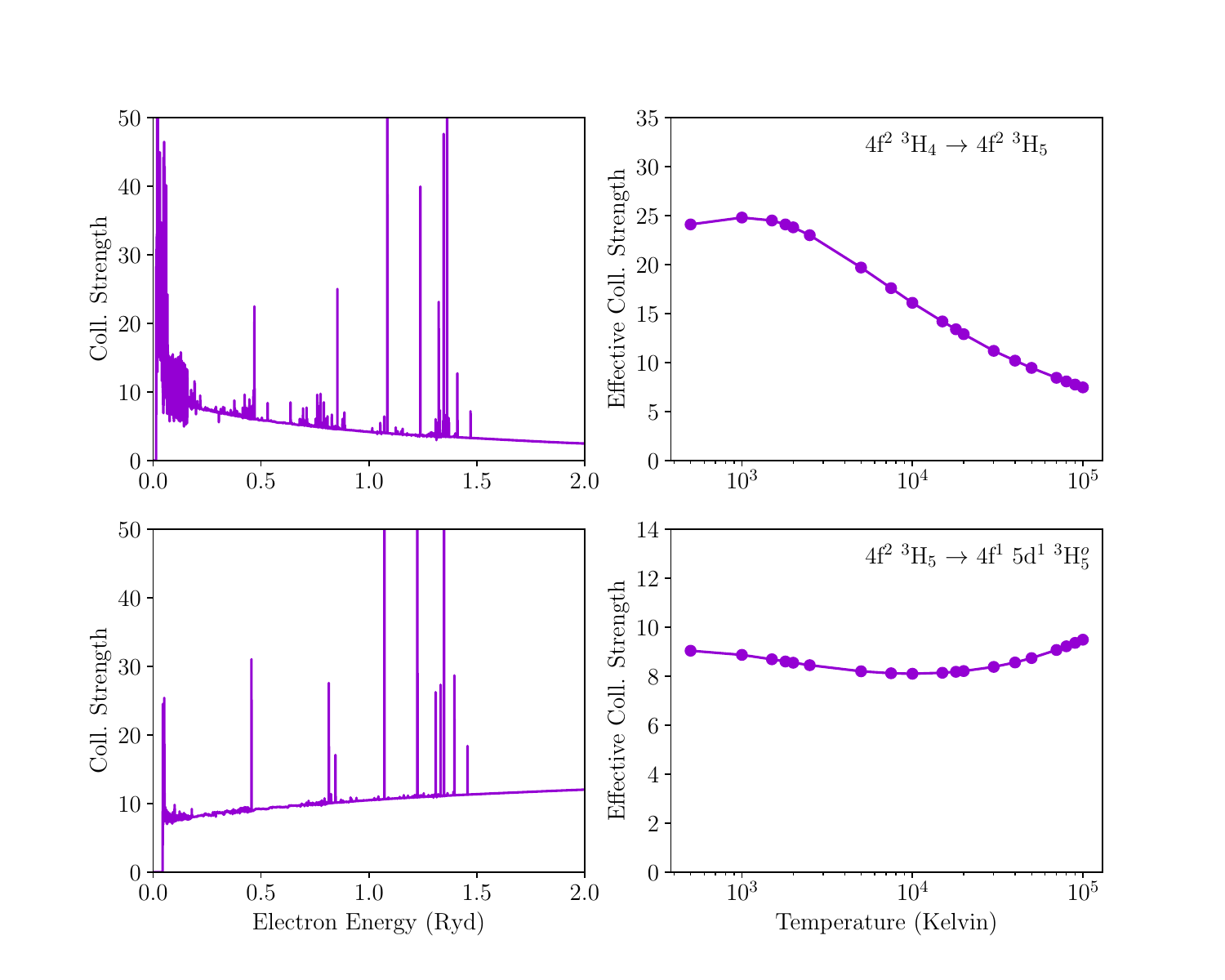}
    \caption{Collision strengths for two transitions of Ce {\sc iii}. The top two panels show the collision strengths and effective collision strengths for the forbidden transition 4f$^2$ $^3$H$_{4}$ $\to$ 4f$^2$ $^3$H$_{5}$ (Level 1 $\to 2$). The bottom two panels show the allowed transition 4f$^2$ $^3$H$_{5}$ $\to$ 4f$^1$ 5d$^1$ $^3$H$^o_{5}$ (Level 2 $\to $12).}
    \label{fig:CeIIICollisions}
\end{figure*}

\begin{figure*}
    \centering
    \includegraphics[width = 0.7\linewidth]{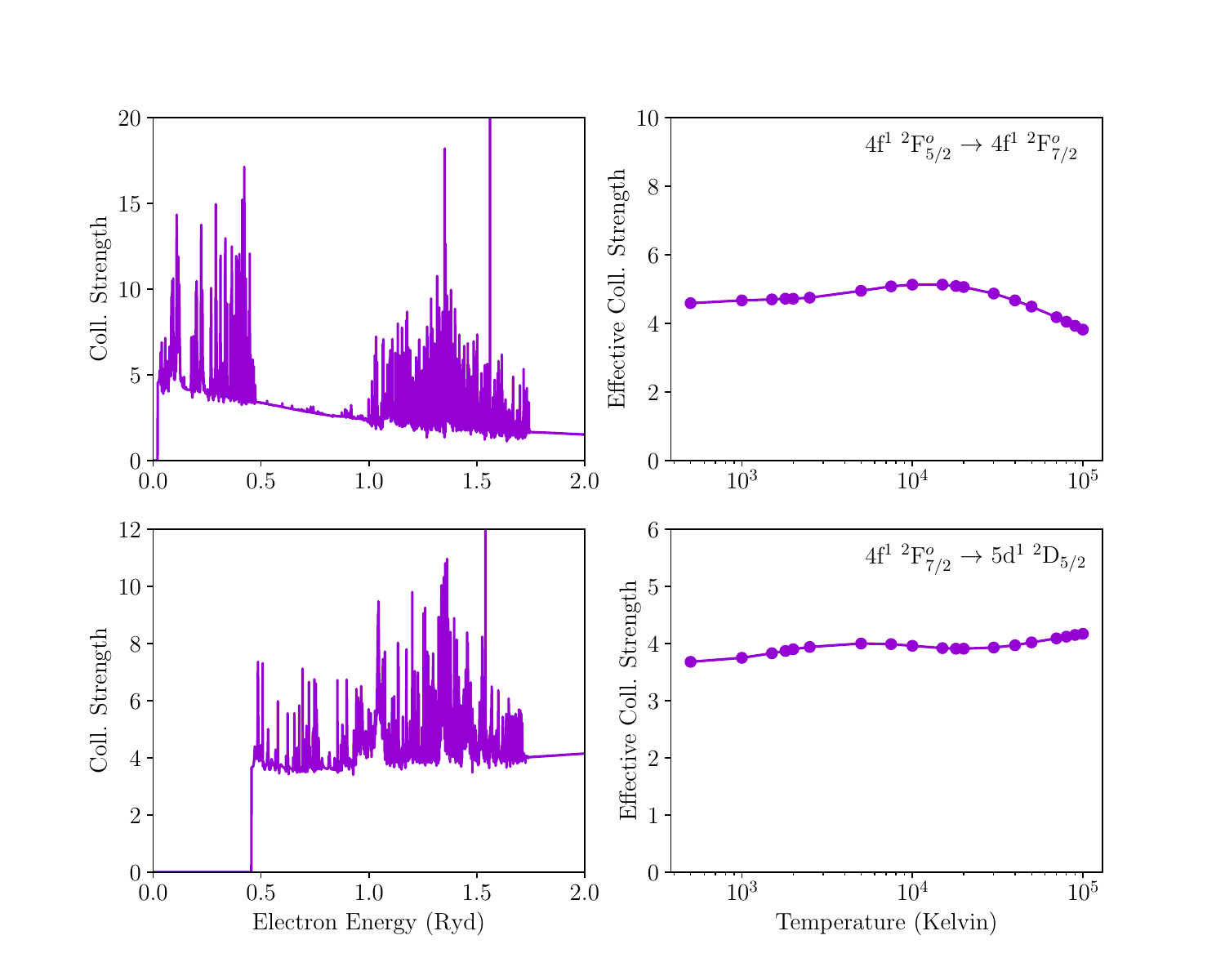}
    \caption{Collision strengths for two transitions of Ce {\sc iv}. The top two panels show the collision strengths and effective collision strengths for the forbidden transition 4f$^1$ $^2$F$_{5/2}^o$ $\to$ 4f$^1$ $^2$F$_{7/2}^o$ (Level 1 $\to 2$). The bottom two panels show the allowed transition 4f$^1$ $^2$F$_{7/2}^o$ $\to$ 5d$^1$ $^2$D$_{5/2}$ (Level 2 $\to $4).}
    \label{fig:CeIVCollisions}
\end{figure*}

We compute collision strengths explicitly for the ions Ce {\sc ii} - {\sc iv}. We detail the main calculation parameters in Table \ref{tab:rmatrixParameters}. In all cases, partial waves not explicitly calculated are accounted for by the sum-rule of \cite{burgess1974coulomb}. The integral in Eq. \eqref{eq:eff} for dipole transitions is supplemented by high energy limits \citep{bethe1930theorie,burgess1992analysis}. In all cases, the first 32 partial waves are calculated on a fine energy mesh detailed in Table \ref{tab:rmatrixParameters}. The remaining partial waves are smoothly varying and are calculated on coarse grid and interpolated onto the fine grid. Figures \ref{fig:CeIICollisions}, \ref{fig:CeIIICollisions} and \ref{fig:CeIVCollisions} show the collision strengths and effective collision strengths for one allowed and one forbidden transition for each of the three systems investigated. The data shown form a subset of transitions of potential importance for cooling and emission modelling of KNe ejecta, and are chosen as typical examples of the behaviour of forbidden and allowed collision strengths. These are then proportional to the excitation rates to be employed in Section \ref{sec:crm}.

\section{Recombination Rate Coefficients} \label{sec:recomb}

In addition to the relativistic R-matrix collision calculation, we also perform a semi-relativistic calculation of the relevant recombination rates using the {\sc autostructure} code \citep{badnell}. This code adopts a Breit-Pauli Hamiltonian, with semi-relativistic radial functions analogous to the HFR method of \cite{cowan1981theory}.

The recombination process is computed perturbatively by means of the distorted wave approach. This separates the direct process (radiative recombination, RR) from the resonant (dielectronic recombination, DR) process. The total rate coefficient is then given by,
\begin{equation}
    \alpha^{\text{Rec}}_{f\to b} =\alpha^{\text{RR}}_{f\to b} + \alpha^{\text{DR}}_{f\to b}, 
\end{equation}
where $f$ is some free state and $b$ some bound state. Radiative recombination is calculated as the statistical inverse of non-resonant photoionization with rate coefficient,
\begin{equation}
    \alpha^{\rm{RR}}_{f \to b} = \frac{4}{\sqrt{2\pi m}}\frac{1}{(kT_e)^{3/2}}\int_0^{\infty} \text{d}\epsilon_f ~e^{-\epsilon_f/kT_e} \epsilon_f~\sigma^{\rm{RR}}_{f\to b} \label{eq:recrate},
\end{equation}
where $\epsilon_f$ is the energy of the captured electron and
\begin{equation}
    \sigma^{\rm{RR}}_{f\to b} = \frac{1}{2} \frac{g_b}{g_f} \frac{ E_p^2}{mc^2\epsilon_f } \sigma^{\rm{DPI}}_{b\to f} \label{eq:detailedBalance},
\end{equation}
where $\sigma^{\rm{DPI}}_{b\to f}$ is the non-resonant photoionization cross section at photon energy $E_p$.

DR proceeds via a two-step process. First, an electron is temporarily captured in the potential of the parent ion (dielectronic capture). This is immediately followed by either a radiative decay into true bound states of the recombined ion or autoionization. Therefore, the branching ratio,
\begin{equation}
    B_{j\to k} = \frac{A^r_{j\to k}}{\sum_{k'}A^r_{j\to k'} + \sum_{f}A^a_{j\to f}},
\end{equation}
gives the fraction of atoms in autoionizing state $j$ that will radiatively stabilize to state $k$. Here, $A^r$ represents radiative rates and $A^a$ autoionization rates. The corresponding cross section is then given by a sum over the resonances $j$. The Maxwellian-averaged rate coefficient is then given by,
\begin{equation}
    \alpha_{f \to b} = \frac{h^3} {(2\pi m_ekT_e)^{3/2}} \sum_j \frac{g_j}{2 g_f} e^{-E_{fj} / kT_e} A^a_{j \to f} B^{\text{DR}}_{j\to b},
\end{equation}
for $g_j$ the statistical weight of level $j$ and $E_{fj}$ the resonance energy.

\begin{figure}
    \centering
    \includegraphics[width=\linewidth]{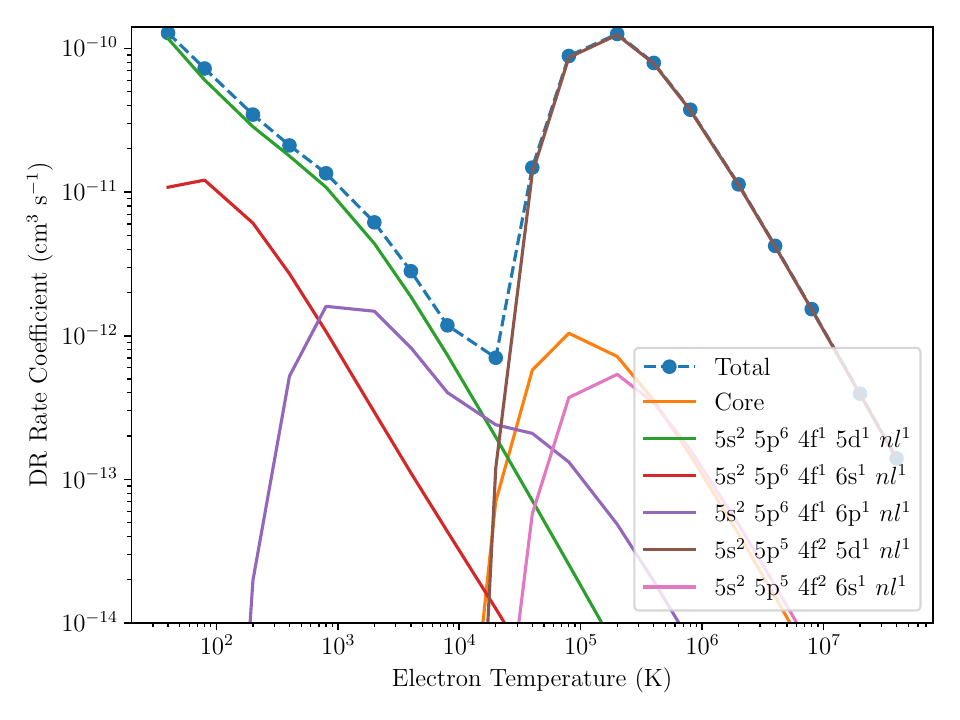}
    \caption{Configuration averaged dielectronic rate coefficients for Ce {\sc iii} $\to$ {\sc ii}. This simplified calculation allows for the identification of the important Rydberg series.}
    \label{fig:configavg}
\end{figure}
In this work, we calculate radiative and dielectronic recombination rate coefficients for Ce {\sc ii} - {\sc vi}. The input configurations were restricted to only contain those configurations that greatly contribute to the resonance pathways. These were initially determined by an inexpensive configuration-averaged calculation to determine the important contributing configurations to guide subsequent level-resolved calculations. As an example, we show the configuration resolved dielectronic recombination rate coefficients on Figure \ref{fig:configavg}. In this case, it is clear that the dielectronic recombination is dominated by the Rydberg series generated by the 5p$^6$4f$^1$ 5d$^1$ and  5p$^5$4f$^2$ 5d$^1$ configurations, giving rise to the low and high temperature behaviour respectively. Similar calculations were carried out for the other ions considered here, finding a similar trend with excited configurations dominating the low temperature, and excitations from the 5p$^6$ core dominating the high temperature. Additional calculation details, including the final set of configurations and determination of radial functions are given in Appendix \ref{sec:rec_appendix}.

\begin{figure*}
    \centering
    \includegraphics[width=\linewidth]{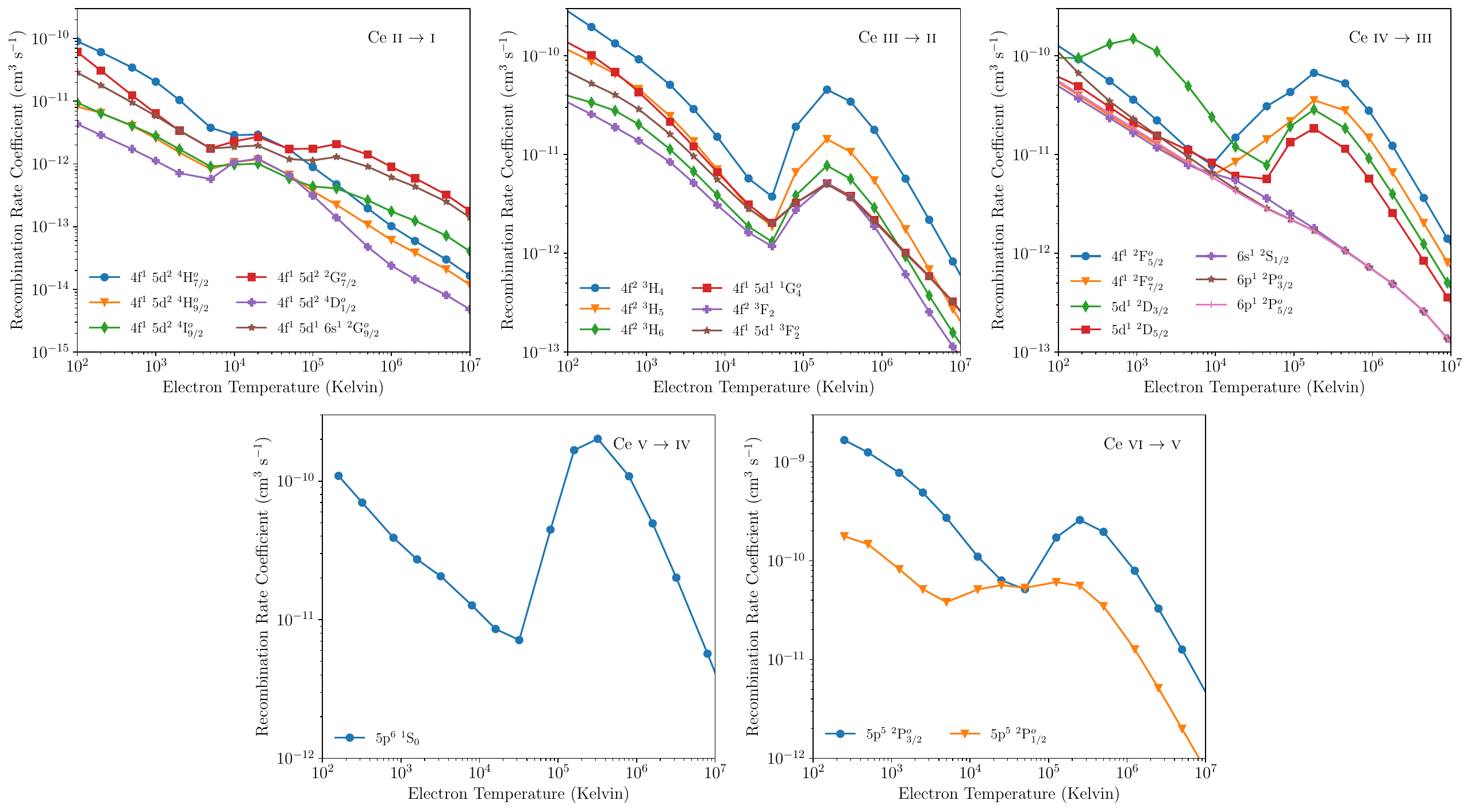}
    \caption{ Total recombination rates for a selection of levels from  Ce {\sc ii} to {\sc vi}. In all cases, the radiative recombination and dielectronic contributions are added together. Additionally, the rates to all of child states are included in the summation.}
    \label{fig:recombinationrate}
\end{figure*}

Using the excitation rates calculated in Sec. \ref{sec:rmatrix}, we were also able to determine which metastables could be significantly populated. As such, we have restricted the recombination calculations to reactions beginning from those levels. The total recombination rates, i.e the sum of the RR and DR components are shown on Figure \ref{fig:recombinationrate} for each of the reactions studied in this work. The rates shown are from a specific metastable of the parent (recombining) ion, to the sum of all of the child (recombined) levels. The recombination rate coefficients are made available.

\section{Modelling} \label{sec:crm}

In this section, we perform collisional radiative modelling with the atomic data calculated in Sections \ref{sec:atomic}, \ref{sec:rmatrix} and \ref{sec:recomb} to investigate the detectability of Ce in nebular kilonovae. 

Assuming a dynamical ejecta mass of around $\sim 5\times 10^{-2} \text{M}_{\odot}$ (see e.g \citealt{waxman2018constraints}, also \citealt{smartt2017kilonova,tanaka2017kilonova}), and a solar $r$-process fraction of $\sim 2\times10^{-3}$ \citep{2009LanB...4B..712L,prantzos2020chemical}\footnote{We take the mass fractions of \cite{2009LanB...4B..712L}, multiplying by the $r$-process mass fraction of \cite{prantzos2020chemical} (see their table 4) and the resulting list is normalized to unit sum.} this results in a rough Ce mass of $\sim  10^{-4} \text{M}_{\odot}$. We refer to this a 'solar' estimation of the Ce mass in KNe. \cite{domoto2022lanthanide} estimate a conservative range of mass fractions of Ce to be $\sim$ $(1-100)\times 10^{-5}$, based on the absorption of allowed Ce {\sc iii} $\sim 1.6 \mu$m. With an ejecta mass of $\sim 0.03\rm{M}_{\odot}$ in their model, this results in $\lesssim 3\times 10^{-5}\rm{M}_{\odot}$ of Ce. The above mass estimations are then in approximate agreement.

To constrain the Ce mass in this work, we employ a one-zone mass model with Ce number density,
\begin{equation}
 n \approx 480 ~\text{cm}^{-3}~  \left( \frac{M_{\text{Ce}}}{10^{-4}\text{M}_\odot}\right), 
\end{equation}
where $M_{\text{Ce}}$ is the mass of Ce in the ejecta in solar masses. We have assumed a homologous expansion of maximum velocity $v=0.1c$, at time $t=29$ days post explosion. The goal of this section is to constrain the parameter $M_{\text{Ce}}$ such that Ce produces emission in similar magnitude to nebular spectra. The collisional radiative equations are solved subject to the constraint that,
\begin{equation}
    n = \sum_{s,i}f_{i}^sn^s,
\end{equation}
where $n^s$ is the number density of ionization state $s$ and $f_{i}^s$ is the fraction density of ionization state $s$ in level $i$. In this section, indices in superscript will refer to an ionization state and those in subscript will refer to levels. Assuming that transitions between ion stages operate on timescales much different to those between levels, we can separate the problem into two components. One for the level populations within an ion stage, and another for the ionization balance.  

\subsection{Level Populations}

The level populations are determined by the steady state of a rate equation,
\begin{equation}
      f^s_i \left(\sum_{j\neq i} n_e q_{i\to j} +\sum_{j < i} A^{\text{eff}}_{i \to j}\right) = \sum_{j\neq i} f^s_j n_e q_{j\to i} + \sum_{j>i}f^s_j A^{\text{eff}}_{j\to i}, \label{eq:crm}
\end{equation}
considering only electron impact excitation/de-excitation and radiative decay. Studies involving a rigorous treatment of the photon field are reserved for more detailed radiative transfer simulations, although we do include the self-absorption of strong lines with, $A^{\text{eff}}_{j\to i} = \beta_{j\to i}~A_{j\to i}$, where $\beta_{j\to i}$ is the escape probability calculated in terms of the \cite{sobolev1957diffusion} optical depth,
\begin{align}
    \tau_{j\to i} &= \frac{\lambda_{j \to i} ^3}{8\pi}g_jA_{j \to i} t ~n^s~\Big(\frac{f^s_i}{g_i}-\frac{f^s_j}{g_j}\Big),\\
    \beta_{j\to i} &= \frac{1}{\tau_{j\to i}} \left(1-e^{-\tau_{j\to i}} \right),
\end{align}
where $t$ is the time since explosion. The system Eq. \eqref{eq:crm} is first solved with $ \beta_{j\to i} = 0$, and solved iteratively with updated values of $ \beta_{j\to i}$ until self consistency. This is particularly important for this case, as the low lying levels of the Ce {\sc ii} and {\sc iii} feature dipole allowed transitions with large reabsorption probabilities. 

The emission of a particular line $j \to i$ from ion stage $k$ is then calculated as,
\begin{equation}
    L^k_{j\to i} = \frac{hc}{\lambda_{ij}} A^{\text{eff}}_{j\to i} f^k_j\frac{n^k}{n}  \frac{M_{\text{Ce}}}{\mathscr{A} m_p},
\end{equation}
where $\mathscr{A} \approx140$ is the nuclear mass of Ce and $m_p$ is the nuclear mass unit. These are then multiplied by a normalized Gaussian kernel, with full-width-half maximum in wavelength space of $\lambda_{ij} v/c$, for $v$ the assumed expansion velocity.  Calculations involving more realistic lineshapes are reserved for future work (see e.g \citealt{jerkstrand17,simotas25,sim2026latetimeemissionlineprofileskilonova}).

\begin{figure*}
    \centering
    \includegraphics[width=\linewidth]{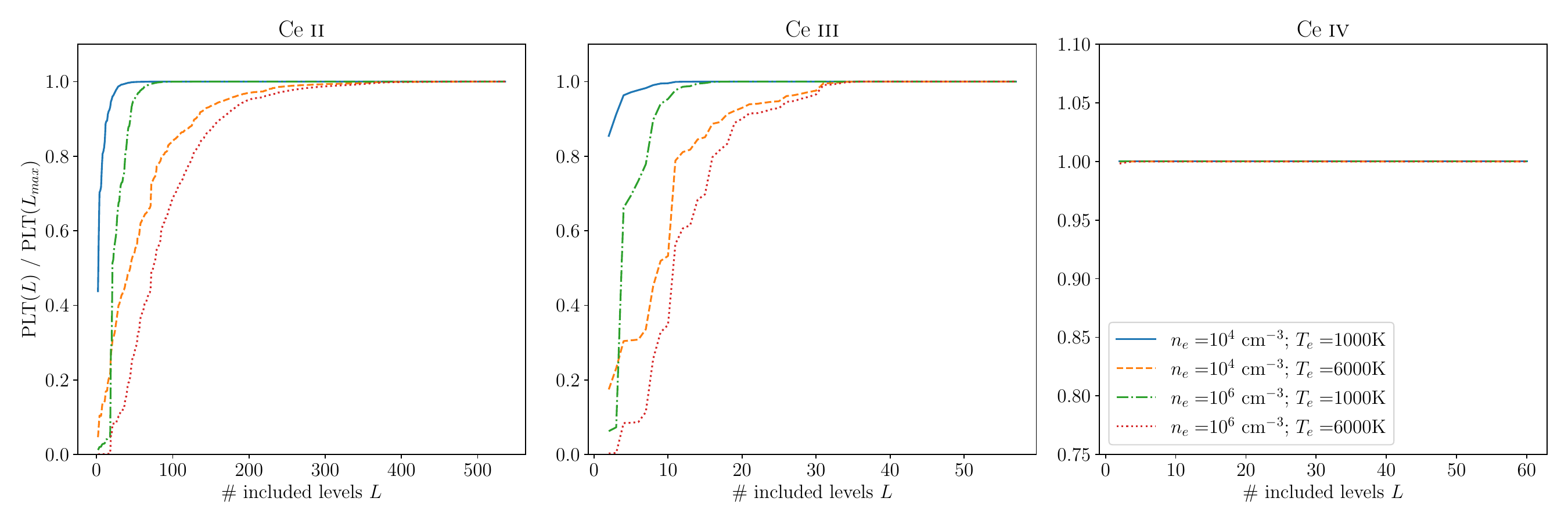}
    \caption{Cooling functions with a truncated expansion of Eq \eqref{eq:crm}. The cooling functions are normalized to those of the maximum number of levels available to this work.}
    \label{fig:pltconvergence}
\end{figure*}

Given the large number of levels involved in the near neutral stages of Ce, we carried out a sensitivity study for the populations as a function of the number of levels included. By varying the number of levels, we effectively truncate the collisional radiative matrix, and therefore test the convergence of the cooling function below,
\begin{equation}
    \text{PLT}^s(T_e,n_e,L) = \sum_{j=2}^L\sum_{i=1}^{j-1} f^s_jA^{\text{eff}}_{j\to i}E_{j\to i},
\end{equation}
where $E_{j\to i}$ is the transition energy. In principle, if the cooling calculated with reduced numbers of levels included in the CR matrix provides a similar result to that of including all available levels at a particular set of plasma parameters - a realistic radiative transfer simulation can truncate the number of levels for this ion and save computation time. The cooling functions normalized to that of the full calculation are shown in Figure \ref{fig:pltconvergence} for $T_e = 1000, 6000$K and $n_e=10^4$, $10^6$ cm$^{-3}$. These parameters represent an approximate range for the for nebular KNe (although more extreme temperatures are reported in models, see e.g \citealt{pognan_steady}). For Ce {\sc ii} it can be seen that in the low temperature regime, the cooling can be safely calculated with $\sim 100$ levels. However, this must be extended to $ \gtrsim 300$ levels at the higher temperatures. Our data set includes  $\sim 550$ levels. Given that matrix inversions scale as $O(L^3)$ - this kind of truncation may save a factor $\sim 10 - 100$ in time spent solving the collisional radiative matrix. By contrast for Ce {\sc iii}, the required number of levels is $\sim 30$ which is much more computationally feasible for current codes. Additionally, most of the cooling for Ce {\sc iv} occurs through the level $2\to1$ transition and the cooling is practically converged when including just this pair of levels. This is primarily because levels higher than the first excited ( $\gtrsim 6.2$ eV) are not typically accessible for this ion under typical KNe conditions.

\subsection{Charge State Balance}

The number densities of each ion stage $n^j$ are estimated by the balance of effective transition rates, 
\begin{equation}
    n^{i+1}n_e\alpha^{i+1\to i} = n^i \left(S^{i\to i+1} + \mathcal{R}^{i\to i+1}\right), \label{eq:balance}
\end{equation}
where  $\alpha^{i+1\to i}$ are recombination rate coefficients, $S^{i\to i+1}$ are the collisional ionization rates and $\mathcal{R}^{i\to i+1}$ are the rates of photoionization by recombination photons (photon recycling). We now discuss each of these contributions.

The recombination rate coefficients are taken as those of the ground-state recombination rates from Section \ref{sec:recomb}.

The collisional ionization rates are obtained from the Spencer-Fano solver {\sc pynonthermal} \citep{luke_shingles_2026_19814117}, which uses semi-empirical electron-impact-ionization cross sections \citep{lotz1968electron,Axelrod1980} and the theory of non-thermal (fast) electron deposition \citep{spencer1954energy, kozma1992gamma}. A detailed study investigating the use of semi-empirical cross sections against distorted-wave cross sections has been carried out for Te ($Z=52$) by \cite{bromley2026atomic}, who find only minor differences in the modelling implications. In the nebular phase of KNe, fast electrons are considered to be the dominant source of ionization (see e.g \citealt{brethauer2026nonthermal}). The Spencer-Fano solver outputs a degradation spectrum $y(\epsilon)$, which combined with the electron impact ionization cross sections gives an ionization rate as,
\begin{equation}
    S^{j\to j+1} = \int \dd \epsilon ~ \sigma^{j\to j+1}(\epsilon) y(\epsilon).
\end{equation}
We aim to gauge the potential impact of Ce on nebular KNe spectra, such as that at $\sim 29$d obtained for AT2023vfi. For this, we have taken the initial deposition rate per ion of fast particles from radioactive decay as $E_{\rm dep} \approx 5\times 10^{-3}~\text{eV}~\text{s}^{-1}$, which was obtained by averaging over the 3D ejecta from the simulations of \cite{shingles2023self}.

In solving Eq. \eqref{eq:balance}, we enforce charge neutrality of a gas of mass $M_{\rm{tot}}$, assuming the entire gas behaves similarly to Ce. That is to say, for $X = M_{\rm{Ce}} / M_{\rm{tot}}$ we have,
\begin{equation}
    n_e = \frac{1}{X}\sum_{k\geq0} kn^k.
\end{equation}
This self-consistently obtained electron density is enforced in the solution of the Spencer-Fano equations, along with a total deposition rate per volume of $n_{\rm{tot}}E_{\rm dep}$ - where $n_{\rm{tot}}$ is the number density corresponding to ejecta mass $M_{\rm{tot}}$ (we assume an average nuclear mass of $\sim 140m_p$).

We additionally considered the potential impact of photoionization by recombination photons. We follow the approximations made in \cite{Axelrod1980} with,
\begin{equation}
    n^i\mathcal{R}^{i\to i+1} =  \sum_{j=i}^{j_{\text{max}}}n^{j+1}n_e\alpha^{j+1\to j}P_{ij}, \label{eq:recycling}
\end{equation}
where $P_{ij}$ is the probability that a recombination into ionization state $j$ releases a photon that photoionizes ionization state $i$.  The summation runs from $i$ to the highest considered charge state $j_{\text{max}}=6+$, and accounts for the potential for recombination between high charge states producing blue photons capable of photoionization. The probabilities $P_{ij}$ are given by,
\begin{equation}
    P_{ij} = \frac{n^i \sigma_i(h\nu_j)}{\sum_k n^k \sigma_k(h\nu_j)} \left(1-e^{-\tau_j}\right)\phi_R,
\end{equation}
for an effective optical depth $\tau_j$ given by 
\begin{equation}
    \tau_j = R\sum_i n^i \sigma_i(h\nu_j),
\end{equation}
where $R= vt\approx 7.5 \times 10^{15} \text{cm}$ is the characteristic length of the line of sight, $\sigma_i$ is the photoionization cross section from state $i$ and $h\nu_i$ is the ground ionization potential of ionization state $i$. Finally, the parameter $\phi_R \leq 1$ is the fraction of the recombination photons that are recycled in the optically thick limit. This quantity being less than unity accounts for the possibility of red recombination photons to excited states that are not energetic enough to cause photoionization (see e.g \citealt{Axelrod1980,hotokezaka2021nebular}).

In solving Eq. \eqref{eq:balance}, an initial (arbitrary in principle) ionization balance is assumed. From this, ionization rates are calculated from the non-thermal deposition solution from {\sc pynonthermal}. The photon recycling rate are then calculated according to Eq. \eqref{eq:recycling} and iterated with Eq. \eqref{eq:balance} to self consistency. These two steps are repeated until full self consistency.

\begin{figure*}
    \centering
    \includegraphics[width=\linewidth]{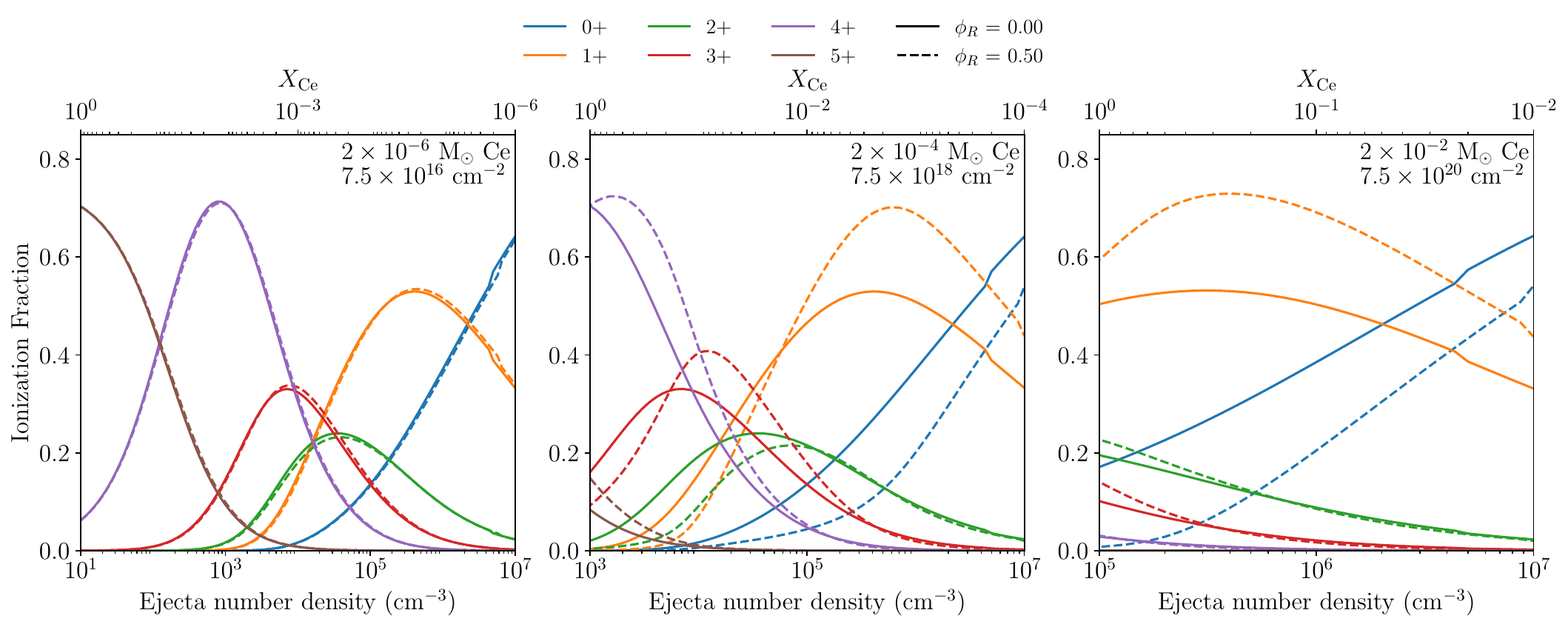}
    \caption{Ionization balance as a function of total number density of the ejecta, at three values of the Ce mass. We additionally indicate the corresponding column densities $Rn$.}
    \label{fig:balance}
\end{figure*}

We show the resulting ionization balance for $\phi_R=0.0$ and $0.5$ as a function of the ejecta density $n_{\rm{tot}}$ in Figure \ref{fig:balance}. We show this at $T_e =3000$K for three masses of Ce, namely $\sim2\times10^{-6,-4,-2}$M$_\odot$, to represent three cases of weak, $\sim$~solar and strong production. These correspond to Ce number densities of $\sim 10^{1,3,5}$ cm$^{-3}$. In the latter two cases, photon recycling suppresses the neutral fraction, and is generally seen to mostly enhance the singly ionized fraction. This is naturally much more pronounced in the higher Ce mass case, as recombination photons will have a higher chance of encountering particles to photoionize at this higher atomic density. Additionally, the optically thin low mass case is unaffected by photon recycling.

We therefore emphasise the importance of detailed radiative transfer models that incorporate this process for definitive interpretations, particularly in models with a realistic density profile in homologous expansion. Photon recycling has been shown to be important in the case of \text{SN~Ia} (see e.g \citealt{Axelrod1980}), and the above shows its effect may be significant for lanthanide rich ejecta.  The use in this work of the scaling parameter $\phi_R$ allows for efficient exploration of parameter space and possible sampling of many photon fields, but is ultimately a simplified model and more detailed treatment is recommended (see e.g \citealt{hotokezaka2021nebular}). Additionally, the large flux of blue recombination photons (e.g Ce {\sc iii} $\to$ Ce {\sc ii} has $h\nu \sim 11$ eV) may raise questions about dust survival in KNe ejecta \citep{hoang2015origin}. Dust is currently an open question in KNe literature \citep{arunachalam2026grb}. \cite{domoto2026heavy} have investigated dust formation in KNe as a way to explain the blackbody-like continuum of AT2023vfi. In particular \cite{domoto2026heavy} have focussed on tungsten $N$-mers, which have a sublimation energy of $\sim 8.9$ eV per atom. However, the low efficiency of photodesorption (see e.g \citealt{lichtman1978photodesorption}) may allow the coexistence of dust with a recombining atomic gas. This said, small $N$-mers may still be susceptible to photodissociation by recombination photons. We emphasise that more work needs to be done in the modelling of dust-gas interactions in future work.

\subsection{Results}

\subsubsection{Ce {\sc iv} mass constraint}

\begin{figure}
    \centering
    \includegraphics[width=\linewidth]{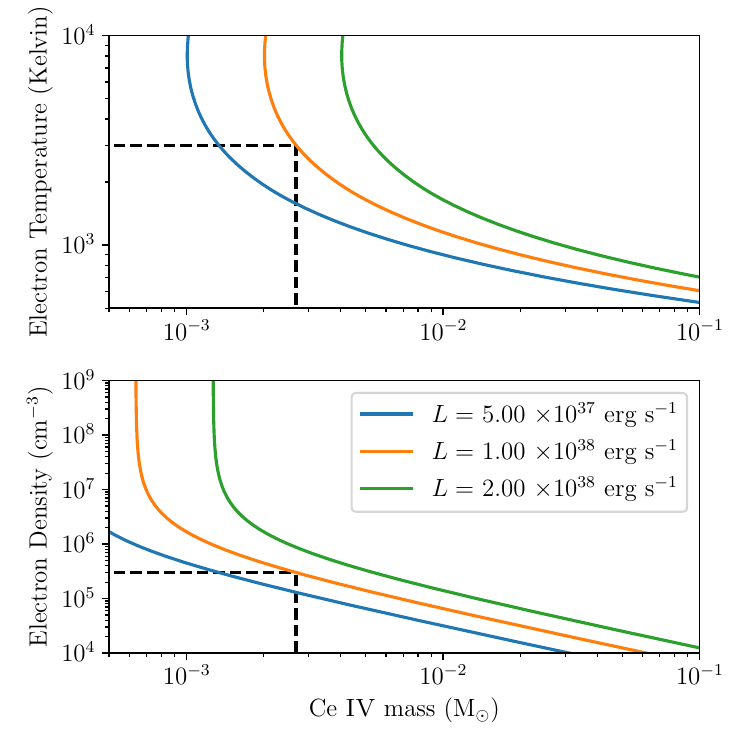}
    \caption{Exploration of mass, density and temperature space - assuming that Ce IV is responsible for the excess flux at $4.5\mu$m in AT2023vfi.}
    \label{fig:contour}
\end{figure}

The low lying line [Ce {\sc iv}] $4438.5$ nm (4f$^1$ $^2$F$_{7/2}^o$ $\to$ 4f$^1$ $^2$F$_{5/2}^o$, 1 to 2) is close to the $\sim 4.5 \mu$m  broad excess flux in AT2023vfi, as well as the photometry of AT2017gfo. Assuming that the entirety of the excess flux ($\sim 10^{38}$ erg s$^{-1}$, see e.g \citealt{gillanders2025analysis}) is due to Ce {\sc iv}, we estimate at the conditions posed at 29d by \cite{levan2024heavy} ($T_e \approx 3000$K, $n_e \approx 3\times 10^5$cm$^{-3}$) that a Ce {\sc iv} mass of $\sim 2.7 \times10^{-3}$ M$_{\odot}$ is required. Figure \ref{fig:contour} shows a rough exploration of the ionic mass required in and around these conditions. It is clear that under all conditions explored here, one needs at least $\sim 10^{-3}$ M$_{\odot}$ to explain this emission in terms of solely Ce {\sc iv}. Furthermore - this will clearly hold true even if the temperature is considerably higher than the limit of 10,000K displayed in Figure \ref{fig:contour}.

It is clear that for this required Ce {\sc iv} mass - a Ce mass fraction much larger than that of the solar pattern is needed. It is noteworthy that this is considerably higher than the amount of Ce estimated by \cite{domoto2022lanthanide} and \cite{gillanders2026improved}, which are both based on the early phases of \gfo. Additionally, it is noteworthy that the excess emission at $\sim 4.5 \mu$m at 29d is not retained in the $\sim$ 60d spectrum (see e.g \citealt{gillanders2025analysis}). Alternatively - if the continuum is explainable by dust \citep{domoto2026heavy}, then the absence of excess flux at $\sim 4.5 \mu$m at 60d may be due to the reduction of dust temperature and the shifting of the blackbody extremum.

\subsubsection{Emission Modelling}

With the goal of investigating the prevalence of Ce in late time emission, it is most appropriate to compare with the 29d post explosion spectrum of AT2023vfi \citep{levan2024heavy,gillanders2023heavy,gillanders2025analysis}. Within this spectrum, notable features appear at $2.1$ $\mu$m and possibly at $4.5$ $\mu$m, that have been discussed in the literature with a variety of interpretations  \citep{Hotokezaka2022WSe,hotokezaka2023tellurium,gillanders2024modelling,Mulholland24,mccann2025luminosity,gillanders2025analysis,jerkstrand2026infrared,pognan2026lanthanide}.  In this section we calculate emission spectra of Ce according to the above prescription. We additionally include a model continuum as that calculated by \cite{gillanders2025analysis}. While the source of the continuum of kilonovae is still debated, it is possible it is produced by emission from lanthanide rich ejecta. None-the-less, we include the continuum in this model as a qualitative notion of whether Ce can individually be detected in future events in the nebular phase. i.e - we are to compare the magnitude of the Ce emission against the magnitude of the model blackbody and afterglow continuum.

\begin{figure*}
    \centering
    \includegraphics[width = \linewidth]{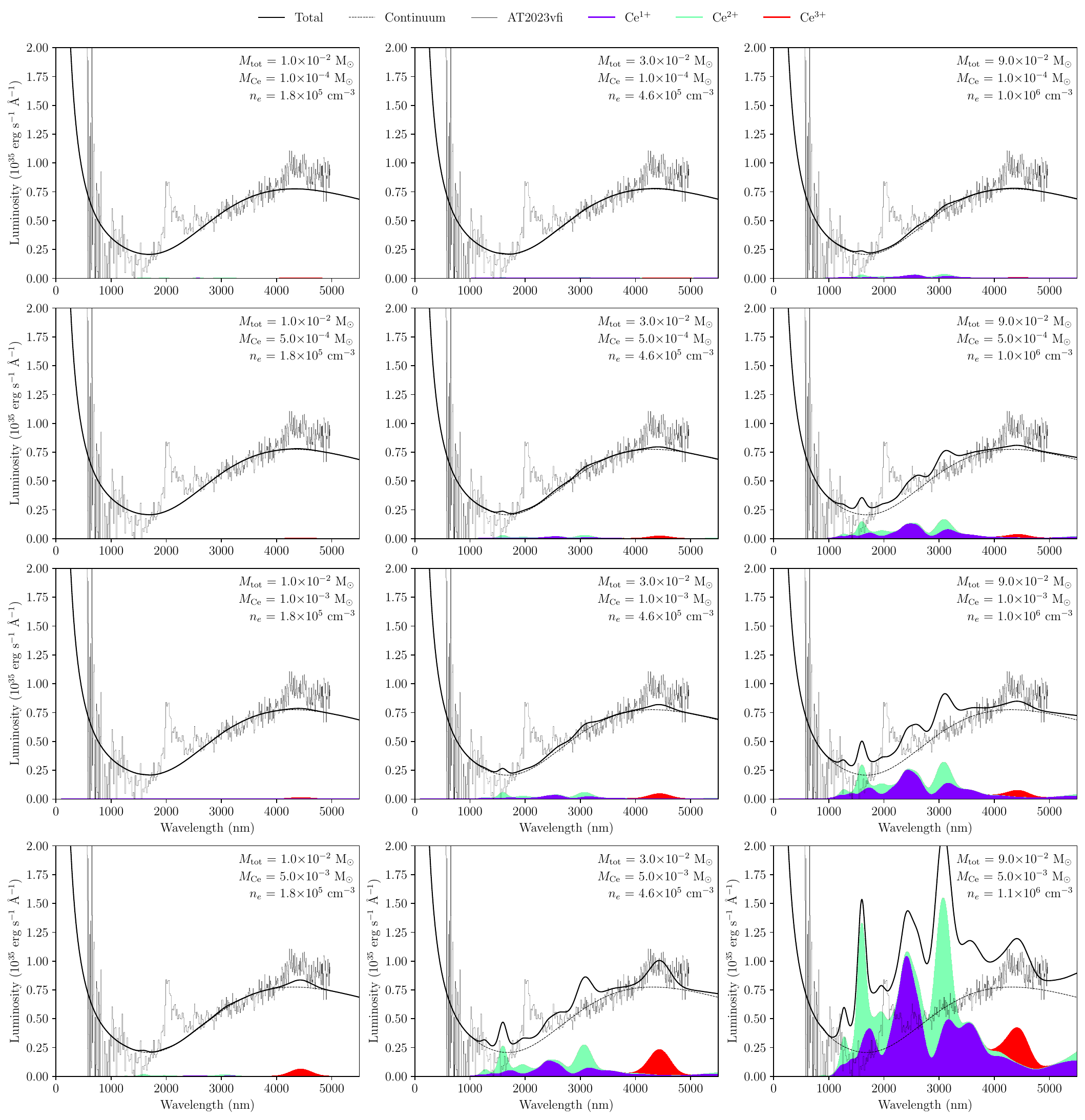}
    \caption{Synthetic spectra calculated with expansion velocity $0.1c$, electron temperature $\approx 3000$K. These are compared with the spectra of AT2023vfi \citep{levan2024heavy}. We use the continuum function of \citet{gillanders2025analysis}. }
    \label{fig:crm}
\end{figure*}

Holding the electron temperature and expansion velocity fixed at $3000$K, $0.1c$ respectively - we calculate synthetic spectra on the grid of total ejecta masses $M_{\rm{tot}} \in \{1, 3,9\} \times 10^{-2}$~M$_{\odot}$,  and Ce masses $M_{\rm{Ce}}\in \{1,5,10,50\}\times 10^{-4}$~M$_{\odot}$. This is shown on Figure \ref{fig:crm}. These spectra are compared with the observed 29d spectrum of AT2023vfi \citep{levan2024heavy}.  At a Ce mass of $1\times 10^{-4}$M$_{\odot}$ ($X\approx 0.002 \sim 1\times$ solar $r$-process), it is seen that Ce does not significantly stand out against the model continuum at the range of electron densities considered. With further increases to the Ce mass, contributions can begin to be seen at higher ejecta density. However, in this regime - the strength of recombination increases and leads to a decreasing average charge state. In particular, it is seen that Ce {\sc iv} gives a very weak contribution, even at high Ce masses. Most notably there are potentially interesting emission features from [Ce II] $\sim$ 2.5 $\mu$m (4f$^2$ 6s$^1$ $^4$H$_{7/2, 9/2}$ $\to$ 4f$^1$ 5d$^2$ $^4$H$_{7/2}^o$, $\lambda = 2.5947, 2.4006$ $\mu$m)  as well as [Ce III] $\sim$ 3.0 $\mu$m (4f$^1$ 5d$^1$ $^1$G$_4^o$ $\to$ 4f$^2$ $^3$H$_4$, $\lambda = 3.0519 \mu$m). These are due to allowed lines which, while modelled with Sobolev opacity here, must be considered more intricately in a true radiative transfer simulation.

The prospect of Ce production as large as those considered in Figure \ref{fig:crm} where Ce begins to stand out from the continuum ($X\gtrsim0.05$) seems unlikely, particularly given the recent re-estimation of the lanthanide fraction of \cite{gillanders2026improved} who find a value much lower than previously estimated. This was primarily due to the new atomic data for opacities calculated by \cite{flors2026calibrated}. This, combined with the overall weakness of Ce in this model - leads one to the notion that it may be extremely difficult to detect lanthanides on an element-by-element basis in the emission dominated parts of kilonovae simulation. This is particularly so given that Ce is close to the second peak of the $r$-process abundance pattern and of even atomic number. I.e if detecting Ce is difficult in emission, then there will be greater difficulty in detecting other lanthanides individually in emission. This outlook is of course constrained to the relatively simple models calculated here, although perhaps points to a future need simply for adequate data to calculate the cooling due to lanthanides as opposed to discrete line identification. The individual detection of lanthanides may be limited to absorption analysis of early time ejecta, such as the studies conducted by \cite{domoto2022lanthanide} and co-workers.

\section{Conclusions and Outlook}\label{sec:conclusions}

In this work we have expanded the atomic data sets for the lanthanide element cerium. We have used the {\sc grasp}$^0$, {\sc darc} and {\sc autostructure} codes to produce atomic structures, collision strengths and recombination data. The calculated Einstein A-coefficients show reasonable agreement with literature values, particularly those of \cite{flors2026calibrated, gaigalas2024theoretical, rynkun2022theoretical}. These, combined with the collision strengths and recombination rates will be made publicly available and will serve to allow the more advanced modelling of Ce in astrophysical application.

By modelling ionization by collision with fast electrons balanced with 
radiative/dielectronic recombination, we have found the recycling of recombination photons into photoionization to be a potentially significant effect, and therefore should be considered in future modelling of lanthanide rich ejecta. Even a basic implementation of this will become more feasible as bespoke recombination calculations are made available as these will also provide the necessary photoionization cross sections. 

In such ionization balances, we have found that the relatively simple lanthanide of Ce with a plethora of strong lines requires significantly strong $r$-process production to individually stand out in the spectra of kilonovae. Given the recent estimations of lanthanide mass fractions by \cite{gillanders2026improved} ($X_{\rm{lan}} \lesssim0.0025$), this is unlikely. Therefore, not only does the correspondence of [Ce {\sc iv}] $\sim 4.5\mu$m with excess flux around this wavelength in \grb~seem unlikely - the detection of Ce in emission in general is potentially infeasible.

It is therefore reasonable to assume that the detection of any of the lanthanides on an individual line-by-line basis may prove infeasible in future spectral analyses of kilonova observations. The detection of lanthanides may then be limited to early time optical depth studies such as those by \cite{domoto2022lanthanide}. This may indicate a shift in the requirement of atomic data by the modelling community. In particular, perhaps what is needed is data of high enough quality to correctly calculate cooling and opacity without the strong requirement of spectral wavelength accuracy. This said, wavelength calibrated studies such as this and \cite{flors2026calibrated} provide data which remains useful for benchmarking, and additionally the modelling of lanthanides in lower velocity environments with smaller line widths.


\section*{Acknowledgements and Funding}\label{sec:acknowledgements}

LPM, LJS, CAR, CPB and SAS are funded/co-funded by the European Union (ERC, HEAVYMETAL, 101071865). Views and opinions expressed are, however, those of the author(s) only and do not necessarily reflect those of the European Union or the European Research Council. Neither the European Union nor the granting authority can be held responsible for them. N.F acknowledges the Science and Technology Facilities Council (STFC) part of the UK Research and Innovation body for their support via studentship.

Thanks to Quentin Pognan and Andreas Flörs for useful discussion. We thank our colleagues at Queen's University Belfast and Auburn University for helpful discussion. 

We acknowledge {\sc numpy} \citep{harris2020array}, {\sc matplotlib} \citep{Hunter:2007} and {\sc scipy} \cite{2020SciPy-NMeth} for data processing and visualization.

We are grateful for use of the computing resources from the Northern Ireland High Performance Computing (NI-HPC) service funded by EPSRC (EP/T022175).  

This work used the DiRAC Data Intensive service (DIaL3) at the University of Leicester, managed by the University of Leicester Research Computing Service on behalf of the STFC DiRAC HPC Facility (www.dirac.ac.uk). The DiRAC service at Leicester was funded by BEIS, UKRI and STFC capital funding and STFC operations grants. DiRAC is part of the UKRI Digital Research Infrastructure. This work used the DiRAC Extreme Scaling service (Tursa) at the University of Edinburgh, managed by the EPCC on behalf of the STFC DiRAC HPC Facility (www.dirac.ac.uk). The DiRAC service at Edinburgh was funded by BEIS, UKRI and STFC capital funding and STFC operations grants. DiRAC is part of the UKRI Digital Research Infrastructure. 

This research used resources of the National Energy Research Scientific Computing Center (NERSC), a Department of Energy Office of Science User Facility using NERSC award FESERCAP0031596.

\section*{Data Availability}

The effective collision strengths and recombination rate coefficients are recorded in the {\sc adas} standard (see e.g  http://open.adas.ac.uk/). The data will be available in this articles online supplementary material and at Zenodo (\url{https://doi.org/10.5281/zenodo.22751056}). Other data underlying this article will be shared on reasonable request to the corresponding author.



\bibliographystyle{mnras}
\bibliography{bibliography} 

\appendix{}

\section{Steps taken to obtain the {\sc grasp$^0$} structures } \label{sec:graspstructure}

The structures discussed in Section \ref{sec:atomic} were obtained by sequentially adding configurations. Some of these configurations have levels assigned to them in the literature, others are included for configuration mixing and others to allow the radial equations within {\sc grasp}$^0$ to converge. In some cases, configurations are removed and certain orbitals held fixed in steps of the calculation. This is detailed in Table \ref{tab:csf_appendix}. All calculations are performed in the Extended-Average-Level mode, where the secular eigenvalue equation is decoupled from the non-linear radial equations. Additionally, all calculations have orbitals up to 4d closed.

\begin{table}
    \centering
    \begin{tabular}{l l }
    \hline \\
    Ce {\sc ii} Model & \\
    Step 1 &   5s$^{2}$ 5p$^{6}$ \{ 4f$^{1}$ 5d$^{2}$, 4f$^{2}$ 5d$^{1}$, 4f$^{3}$  \}\\ 
           & Optimize all orbitals.\vspace{1mm}  \\ 
    Step 2 & Configs from step 1                 \\ 
           & + 5s$^{2}$ 5p$^{6}$ \{4f$^{1}$ 5d$^{1}$ 6s$^{1}$, 4f$^{2}$ 6s$^{1}$,5d$^{3}$ \} \\
           & Optimize 6s. \vspace{1mm}  \\ 
    Step 3 & Configs from step 2        \\ 
           & + 5s$^{2}$ 5p$^{6}$ 4f$^{1}$\{ 5d$^{1}$ 6p$^{1}$,6s$^{1}$ 6p$^{1}$, 6s$^{2}$\} \\
           & + 5s$^{2}$ 5p$^{6}$ 4f$^{2}$ 6p$^{1}$ \\
           & Optimize 6p. \vspace{1mm} \\ 
    Step 4 & Configs from step 3 \\     
           & 5s$^{2}$ 5p$^{5}$ 4f$^{3}$ 6s$^{1}$ \\
           & 5s$^{2}$ 5p$^{6}$ 5d$^{1}$ \{  6s$^{2}$, 6p$^{2}$ \} \\
           & 5s$^{2}$ 5p$^{6}$ 5d$^{2}$ \{6s$^{1}$, 6p$^{1}$\} \\
           & 5s$^{2}$ 5p$^{6}$ \{ 6p$^{3}$,6s$^{1}$ 6p$^{2}$,4f$^{1}$ 6p$^{2}$ \}\\
           & All orbitals fixed. \vspace{2mm}\\
    Ce {\sc iii} Model &                                                    \\ 
    Step 1 & 5s$^{2}$ 5p$^{6}$\{ 4f$^{2}$, 5d$^{2}$\}                       \\
           & Optimize all orbitals.\vspace{1mm}                             \\
    Step 2 & Configs from step 1                                            \\
           & + 5s$^2$ 5p$^6$ 5d$^1$\{ 6s$^1$, 6p$^1$, 6d$^1$  \}           \\
           & + 5s$^2$ 5p$^6$ 4f$^1$\{ 5d$^1$, 6s$^1$, 6p$^1$, 6d$^1$  \}   \\
           & + 5s$^2$ 5p$^5$ 4f$^3$; 5s$^2$ 5p$^4$ 4f$^4$;                  \\
           & Optimize 6s,6p,6d.            \vspace{1mm}                     \\
    Step 3 & Configs from step 2                                            \\
           & + 5s$^{2}$ 5p$^{4}$ \{ 4f$^{1}$ 5d$^{3}$, 4f$^{2}$ 5d$^{2}$\}  \\
           & + 5s$^{2}$ 5p$^{5}$ 4f$^{1}$ 5d$^{2}$                          \\ 
           & Optimize all orbitals. \vspace{1mm}                            \\
    Step 4 & Configs from step 3                                            \\ 
           & + 5s$^{2}$ 5p$^{5}$ 4f$^{2}$ 6p$^{1}$                          \\ 
           & - 5s$^{2}$ 5p$^{4}$ \{ 4f$^{1}$ 5d$^{3}$, 4f$^{2}$ 5d$^{2}$\}  \\ 
           & All orbitals fixed.  \vspace{2mm}                              \\
    Ce {\sc iv} Model & \\
    Step 1& \hspace{2.5mm}5s$^2$ 5p$^6$\{4f$^1$, 5d$^1$, 6s$^1$, 6p$^1$, 6d$^1$, 7s$^1$, 7p$^1$, 7d$^1$, 8s$^1$\} \\
          & + 5s$^2$ 5p$^5$\{4f$^{2}$, 5d$^{2}$, 6p$^{1}$ 7p$^{1}$, 6d$^{2}$, 7s$^{2}$ \} \\ 
          & + 5s$^2$ 5p$^5$ 4f$^{1}$\{5d$^{1}$, 6s$^{1}$, 6p$^{1}$\} \\ 

          & Optimize all orbitals. \\
    Step 2& Configs from step 1 \\ 
          & + 5s$^{2}$ 5p$^{6}$ 5f$^{1}$ \\ 
          & + 5s$^{2}$ 5p$^{5}$ \{4f$^{1}$ 5f$^{1}$, 5f$^{2}$ \} \\ 
          & + 5s$^{2}$ 5p$^{4}$ 5f$^{3}$ \\ 
          & + 5s$^{2}$ 5p$^{3}$ 5f$^{4}$ \\ 
          & Optimize 4f and 5f. \\ 
    Step 3& Configs from step 2 \\ 
          & + 5s$^{0}$ 5p$^{6}$ 4f$^{3}$  \\ 
          & + 5s$^{1}$ 5p$^{6}$ \{4f$^{1}$ 6s$^{1}$, 4f$^{1}$ 5d$^{1}$\} \\
          & + 5s$^{2}$ 5p$^{4}$ 4f$^{3}$\\
          & - 5s$^{2}$ 5p$^{4}$ 5f$^{3}$\\
          & - 5s$^{2}$ 5p$^{3}$ 5f$^{4}$\\
          & All orbitals fixed. \\
    \hline\\
    \end{tabular}
    \caption{Order of additional and removals of electronic configurations in generated the atomic structure models. }
    \label{tab:csf_appendix}
\end{table}

\section{Extended details of recombination rate calculations}\label{sec:appendix_rec} \label{sec:rec_appendix}

Here we show the configuration lists for the recombination reactions detailed in this work, as well as the {\sc autostructure} radial scaling parameters. 

In the generation of child configurations, a Rydberg electron $nl$ is attached to each of the aprent configurations. In all cases, the Rydberg $nl$ is calculated explicitly for DR for $0\leq l\leq6$ and $7\leq n \leq 50$, with additional $n$ values up to $999$ included for interpolation. The RR calculation treats the Rydberg electron similarly, although we restricted its angular momentum to $0\leq l\leq3$, and $l\geq4$ is post-processed with hydrogenic rates, as recommended by the {\sc autostructure} manual \citep{badnell}.

The scaling parameters parameters for the Thomas-Fermi radial functions were optimized using a standard \cite{nelder1965simplex} routine from the {\sc scipy } library \citep{2020SciPy-NMeth}. The optimization was such that the ground levels of both the charge states in each reaction was correct, and additionally that the ionization potentiasl was close to the standard values. This required the additional least-squares optimization of those levels that are close to the ground in either case, and can be erroneously indicated as the ground state in calculations with reduced basis sets employing a strict variational principle. The levels optimized for each reaction are shown in Table \ref{tab:levels_optimized}, along with their recommended energies from the {\sc nist} database, and the final optimized values. Note that we indicate the levels relative to the ground of the recombined charge states, i.e continuum energies include the ionization potential of the recombined ion. In Table \ref{tab:csf_appendix_autostructure} we show the configurations included in these calculations. In Table \ref{tab:radial_parmaters}, the obtained values of the radial parameters are shown. 

Note that the {\sc autostructure} calculations were performed in the relativistic radial method, {\sc icr}, with variable {\sc irel} = 1. This particular method and parameter choice mimics that of the \cite{cowan1981theory} code, where the small radial function is included in the radial equations, but is disregarded with the large component renormalized in the construction of the Hamiltonian. Additionally, a minor modification was made to {\sc autostructure} to allow the use of the same Thomas-Fermi potential for both the core (non-Rydberg autoionizing states) and Rydberg contributions to each process. This allowed for streamlined post-processing and self consistent datasets.

\sisetup{round-mode=places, round-precision=6,group-digits=false}

\begin{table}
    \centering
    \begin{tabular}{l S[table-format=1.6] S[table-format=1.6]}
    \toprule
    Level optimized & {E$_{\text{{\sc nist}}}$} & {E$_{\text{{opt}}}$}\\
    \midrule
    \\
    Ce {\sc ii} $\to $ {\sc i} & & \\
      5p$^6$ 4f$^1$ 5d$^1$ 6s$^2$ $^1$G$^o_4$   & 0.00000    & 0.00000 \\
      5p$^6$ 4f$^1$ 5d$^2$ 6s$^1$ $^5$H$^o_3$   & 0.02158854 & 0.01960400 \\
      5p$^6$ 4f$^2$ 6s$^2$ $^3$H$_4$   & 0.04340108 & 0.05892900 \\
      5p$^6$ 4f$^1$ 5d$^2$ $^4$H$_{7/2}^o$   & 0.40708000 & 0.41524300 \\
      5p$^6$ 4f$^1$ 5d$^1$ 6s$^1$ $^4$F$_{3/2}^o$   & 0.42878862 & 0.42707600 \\
      5p$^6$ 4f$^2$ 6s$^1$ $^4$H$_{7/2}$   & 0.44220034 & 0.42644000 \\
\\
    Ce {\sc iii} $\to $ {\sc ii} & & \\
      5p$^6$ 4f$^1$ 5d$^2$ $^4$H$_{7/2}^o$   & 0.00000    & 0.00000 \\
      5p$^6$ 4f$^1$ 5d$^1$ 6s$^1$ $^2$G$_{9/2}^o$   & 0.02170862 & 0.02170100 \\
      5p$^6$ 4f$^2$  $^3$F$_{4}$   & 0.80530000 & 0.80530000 \\
      5p$^6$ 4f$^1$ 5d$^1$ $^1$G$^o_{4}$   & 0.83515910 & 0.83514700 \\
\\
    Ce {\sc iv} $\to $ {\sc iii} & & \\
      5p$^6$ 4f$^2$  $^3$F$_{4}$   & 0.00000    & 0.00000 \\
      5p$^6$ 4f$^1$ 5d$^1$ $^1$G$^o_{4}$   & 0.0298591 & 0.02985800 \\
      5p$^6$ 4f$^1$  6s$^1$ $^3$F$^o_{2}$   & 0.1752934    & 0.17529200 \\

      5p$^6$ 4f$^1$  $^2$F$^o_{5/2}$   & 1.4844800 & 1.48448200 \\
      5p$^6$ 5d$^1$  $^2$D$_{3/2}$   & 1.9377200 & 1.93771900 \\
\\
    Ce {\sc v} $\to $ {\sc iv} & & \\
      5p$^6$ 4f$^1$  $^2$F$^o_{5/2}$          & 0.00000    & 0.00000 \\
      5p$^6$ 4f$^1$ $^2$F$^o_{7/2}$  & 0.02053 & 0.01875500 \\
      5p$^6$ 5d$^1$ $^2$D$_{3/2}$  &  &  \\

      5p$^6$ 6s$^1$ $^2$S$_{1/2}$  & 0.78918 & 0.78459700 \\
      5p$^6$ 6p$^1$ $^2$P$^o_{1/2}$  & 1.11708 & 1.09770800 \\

      5p$^6$  $^1$S$_{0}$   & 2.7126 & 2.70479000 \\
      5p$^5$ 4f$^1$  $^3$D$_{1}$   & 3.69111728 & 3.69681500 \\
      5p$^5$ 5d$^1$  $^3$P$^o_{0}$   & 4.33777502 & 4.34897300 \\
      \bottomrule
    \end{tabular}
    \caption{Levels that the {\sc autostructure} radial orbitals were optimized against. For each reaction, the levels are given in units of Rydbergs, relative to the ground-state of the \emph{recombined} system. Each of the levels take recommended values from the {\sc nist} database \citep{nist}, except for those of Ce {\sc v}, from which we quote levels from \citet{wajid2021spectral}.}
    \label{tab:levels_optimized}
\end{table}

\begin{table}
    \centering
    \begin{tabular}{l l }
    \toprule 
    Master configurations & Rydberg configurations \\ 
    \midrule
    {Ce {\sc ii} $\to$ Ce {\sc i}} \\ 
5s$^{2}$ 5p$^{6}$ 4f$^{1}$ 5d$^{1}$ 6s$^{2}$  &5s$^{2}$ 5p$^{6}$ 4f$^{3}$ $nl^{1}$ \\
5s$^{2}$ 5p$^{6}$ 4f$^{1}$ 5d$^{2}$ 6s$^{1}$  &5s$^{2}$ 5p$^{6}$ 4f$^{2}$ 5d$^{1}$ $nl^{1}$ \\
5s$^{2}$ 5p$^{6}$ 4f$^{2}$ 5d$^{1}$ 6s$^{1}$  &5s$^{2}$ 5p$^{6}$ 4f$^{2}$ 6s$^{1}$ $nl^{1}$ \\
5s$^{2}$ 5p$^{6}$ 4f$^{2}$ 6s$^{2}$  &5s$^{2}$ 5p$^{6}$ 4f$^{2}$ 6p$^{1}$ $nl^{1}$ \\
5s$^{2}$ 5p$^{6}$ 4f$^{1}$ 5d$^{1}$ 6s$^{1}$ 6p$^{1}$  &5s$^{2}$ 5p$^{6}$ 4f$^{1}$ 5d$^{2}$ $nl^{1}$ \\
5s$^{2}$ 5p$^{6}$ 4f$^{2}$ 5d$^{2}$  &5s$^{2}$ 5p$^{6}$ 4f$^{1}$ 5d$^{1}$ 6s$^{1}$ $nl^{1}$ \\
5s$^{2}$ 5p$^{6}$ 4f$^{3}$ 5d$^{1}$  &5s$^{2}$ 5p$^{6}$ 4f$^{1}$ 5d$^{1}$ 6p$^{1}$ $nl^{1}$ \\
5s$^{2}$ 5p$^{6}$ 5d$^{1}$ 6s$^{2}$ 6p$^{1}$  &5s$^{2}$ 5p$^{6}$ 5d$^{3}$ $nl^{1}$ \\
& 5s$^{2}$ 5p$^{6}$ 5d$^{2}$ 6s$^{1}$ $nl^{1}$ \\
& 5s$^{2}$ 5p$^{6}$ 5d$^{2}$ 6p$^{1}$ $nl^{1}$ \\
& 5s$^{2}$ 5p$^{6}$ 4f$^{1}$ 6s$^{2}$ $nl^{1}$ \\
& 5s$^{2}$ 5p$^{6}$ 5d$^{1}$ 6s$^{2}$ $nl^{1}$ \\
& 5s$^{2}$ 5p$^{6}$ 6s$^{2}$ 6p$^{1}$ $nl^{1}$ \\
& 5s$^{2}$ 5p$^{5}$ 4f$^{1}$ 5d$^{2}$ 6s$^{1}$ $nl^{1}$ \\
\\
{Ce {\sc iii} $\to$ Ce {\sc ii}} \\ 
5s$^{2}$ 5p$^{6}$ 4f$^{3}$ & 5s$^{2}$ 5p$^{6}$ 4f$^{2}$ $nl^{1}$ \\
5s$^{2}$ 5p$^{6}$ 4f$^{2}$ 5d$^{1}$ & 5s$^{2}$ 5p$^{6}$ 4f$^{1}$ 5d$^{1}$ $nl^{1}$ \\
5s$^{2}$ 5p$^{6}$ 4f$^{2}$ 6s$^{1}$ & 5s$^{2}$ 5p$^{6}$ 4f$^{1}$ 6s$^{1}$ $nl^{1}$ \\
5s$^{2}$ 5p$^{6}$ 4f$^{2}$ 6p$^{1}$ & 5s$^{2}$ 5p$^{6}$ 4f$^{1}$ 6p$^{1}$ $nl^{1}$ \\
5s$^{2}$ 5p$^{6}$ 4f$^{1}$ 5d$^{2}$ & 5s$^{2}$ 5p$^{6}$ 5d$^{2}$ $nl^{1}$ \\
5s$^{2}$ 5p$^{6}$ 4f$^{1}$ 5d$^{1}$ 6s$^{1}$ & 5s$^{2}$ 5p$^{6}$ 5d$^{1}$ 6s$^{1}$ $nl^{1}$ \\
5s$^{2}$ 5p$^{6}$ 4f$^{1}$ 5d$^{1}$ 6p$^{1}$ & 5s$^{2}$ 5p$^{6}$ 5d$^{1}$ 6p$^{1}$ $nl^{1}$ \\
5s$^{2}$ 5p$^{6}$ 4f$^{1}$ 6s$^{2}$ & 5s$^{2}$ 5p$^{5}$ 4f$^{2}$ 5d$^{1}$ $nl^{1}$ \\
\\
{Ce {\sc iv} $\to$ Ce {\sc iii}} \\ 
5s$^{2}$ 5p$^{6}$ 4f$^{2}$  & 5s$^{2}$ 5p$^{6}$ 4f$^{1}$ $nl^{1}$ \\
5s$^{2}$ 5p$^{6}$ 4f$^{1}$ 5d$^{1}$  & 5s$^{2}$ 5p$^{6}$ 5d$^{1}$ $nl^{1}$ \\
5s$^{2}$ 5p$^{6}$ 4f$^{1}$ 6s$^{1}$  & 5s$^{2}$ 5p$^{6}$ 6s$^{1}$ $nl^{1}$ \\
5s$^{2}$ 5p$^{6}$ 4f$^{1}$ 6p$^{1}$  & 5s$^{2}$ 5p$^{6}$ 6p$^{1}$ $nl^{1}$ \\
5s$^{2}$ 5p$^{6}$ 5d$^{2}$  & 5s$^{2}$ 5p$^{5}$ 4f$^{1}$ 5d$^{1}$ $nl^{1}$ \\
5s$^{2}$ 5p$^{5}$ 4f$^{3}$ & \\
5s$^{2}$ 5p$^{5}$ 4f$^{2}$ 5d$^{1}$ & \\
\\
{Ce {\sc v} $\to$ Ce {\sc iv}} \\ 
5s$^{2}$ 5p$^{6}$ 4f$^{1}$  & 5s$^{2}$ 5p$^{6}$ $nl^{1}$ \\
5s$^{2}$ 5p$^{6}$ 5d$^{1}$  & 5s$^{2}$ 5p$^{5}$ 4f$^{1}$ $nl^{1}$ \\
5s$^{2}$ 5p$^{6}$ 6s$^{1}$  & 5s$^{2}$ 5p$^{5}$ 5d$^{1}$ $nl^{1}$ \\
5s$^{2}$ 5p$^{6}$ 6p$^{1}$  & 5s$^{2}$ 5p$^{5}$ 6s$^{1}$ $nl^{1}$ \\
5s$^{2}$ 5p$^{4}$ 4f$^{3}$  & 5s$^{2}$ 5p$^{5}$ 6p$^{1}$ $nl^{1}$ \\
\\
{Ce {\sc vi} $\to$ Ce {\sc v}} \\
5s$^{2}$ 5p$^{6}$ & 5s$^{2}$ 5p$^{5}$ $nl^{1}$ \\
& 5s$^{1}$ 5p$^{6}$ $nl^{1}$ \\
& 5s$^{2}$ 5p$^{4}$ 4f$^{1}$ $nl^{1}$ \\
& 5s$^{2}$ 5p$^{4}$ 5d$^{1}$ $nl^{1}$ \\
& 5s$^{1}$ 5p$^{5}$ 4f$^{1}$ $nl^{1}$ \\

    \hline\\
    \end{tabular}
    \caption{Configurations including in the {\sc autostructure} calculations. The left hand column has the so-called 'master' configurations of the recombined ion. The right are the Rydberg autoionizing configurations of the recombined ions. The parent (recombining) ion configurations are obtained by removing the Rydberg electron.}
    \label{tab:csf_appendix_autostructure}
\end{table}

\sisetup{round-mode=places, round-precision=4}
\begin{table}
    \centering
    \begin{tabular}{l  *{5}{S[table-format=1.3, table-align-text-post=true]}}
        \toprule
      Orbital& {\sc ii} $\to $ {\sc i} & {\sc iii} $\to $ {\sc ii} & {\sc iv} $\to $ {\sc iii} & {\sc v} $\to $ {\sc iv} & {\sc vi} $\to $ {\sc v}\\
      \midrule
       1s  & 1.34837625 & 1.36748442 & 0.84162542 & 1.26519675 & 1.29736241 \\
       2s  & 0.83681409 & 0.97835330 & 1.22320591 & 1.15002641 & 1.18327693 \\
       2p  & 0.90458172 & 1.20057279 & 1.15771654 & 1.10360631 & 1.12500448 \\
       3s  & 1.31001759 & 0.97505068 & 0.95287143 & 1.03703450 & 1.04474684 \\
       3p  & 1.17746067 & 1.03972339 & 0.91058096 & 1.03381472 & 1.03538855 \\
       3d  & 0.86327195 & 0.84996765 & 0.76490435 & 1.00343742 & 1.01127591 \\
       4s  & 1.17134059 & 1.07922464 & 0.98895403 & 1.07063210 & 1.08310238 \\
       4p  & 1.01440744 & 1.02865996 & 1.39417787 & 1.06505988 & 1.04983100 \\
       4d  & 1.02321795 & 1.05309321 & 0.99970163 & 1.04471781 & 1.04254318 \\
       5s  & 1.16143323 & 1.10947914 & 1.02753505 & 1.04980355 & 1.08112017 \\
       5p  & 1.07001309 & 1.05308635 & 1.09245110 & 1.05577259 & 1.05643391 \\
       4f  & 1.07855949 & 1.06428231 & 1.02994873 & 1.05865987 & 1.01818471 \\
       5d  & 1.08751194 & 1.06629832 & 0.96120526 & 1.13478729 & 1.09209131 \\
       6s  & 0.97349196 & 1.00087098 & 1.07777594 & 1.02285229 & \\
       6p  & 0.98879820 & 1.00517171 & 1.02422756 & 1.09050672 & \\
       \bottomrule
    \end{tabular}
    \caption{Optimum scaling parameters produced from the energy functional set by the levels in Table \ref{tab:levels_optimized}.}
    \label{tab:radial_parmaters}
\end{table}





\bsp	
\label{lastpage}
\end{document}